\documentclass{article}
\usepackage{fullpage}
\usepackage{times}
\usepackage{amssymb}
\usepackage{graphicx}
\usepackage{amsmath}
\usepackage{amsthm}
\usepackage{subfigure}
\usepackage{bigints}

\theoremstyle{plain}
\theoremstyle{plain}
\theoremstyle{definition}

\begin{document}

\title {General Dynamics of Varying-Alpha Universes}
\author{John D. Barrow\footnote{Email: J.D.Barrow@damtp.cam.ac.uk}\ {}\ and Alexander A. H. Graham\footnote{Email: A.A.H.Graham@damtp.cam.ac.uk}\\
Department of Applied Mathematics and Theoretical Physics\\
Centre for Mathematical Sciences\\
University of Cambridge\\
Wilberforce Road, CB3 0WA, UK}
\date{\today}
\maketitle

\begin{abstract}
We introduce and study extensions of the varying alpha theory of Bekenstein-Sandvik-Barrow-Magueijo to allow for an arbitrary coupling function and self-interaction potential term in the theory. We study the full evolution equations without assuming that variations in alpha have a negligible effect on the expansion scale factor and the matter density evolution, as was assumed in earlier studies. The background FRW cosmology of this model in the cases of zero and non-zero spatial curvature is studied in detail, using dynamical systems techniques, for a wide class of potentials and coupling functions. All the asymptotic behaviours are found, together with some new solutions. We study the cases where the electromagnetic parameter, zeta, is positive and negative, corresponding to magnetic and electrostatic energy domination in the non-relativistic matter. In particular, we investigate the cases where the scalar field driving alpha variations has exponential and power-law self-interaction potentials and the behaviour of theories where the coupling constant between matter and alpha variations is no longer a constant.

\end{abstract}

\section{Introduction}
The fine structure constant, defined in cgs units by $\alpha =e^{2}/\hbar {c}$, is believed to be one of the fundamental constants of nature, governing the strength of electromagnetic interactions below the electroweak scale. Despite its special status, there is a history of theories in which $\alpha $ is allowed to vary slowly in space and time. Historically, the original motivation was Gamow's attempt in 1967 \cite{gamow} to rescue Dirac's proposal \cite{dirac37, dirac38} to introduce a varying gravitation constant, $G\propto t^{-1}$ to explain the large number coincidences of cosmology. Varying $G\propto t^{-1}$ produced dire consequences for the Earth's climate history -- boiling oceans in the pre-Cambrian era \cite{teller48} because the surface temperature of the Earth then varied as $T_{e}\propto t^{-9/4}$ -- and so Gamow proposed replacing it by a time variation in $e^{2}\propto t$ which did not affect the dynamics of the solar system and created a milder thermal history. Teller had also argued that the numerical coincidence $\alpha ^{-1}\simeq \ln (hc/Gm_{pr}^{2})$ suggested that $\alpha $ might fall logarithmically with time if Dirac's arguments were believed and Stanyukovich had also considered varying $\alpha $ in earlier work \cite{stan}; for a review of this early work see \cite{BT}. Gamow's fast variation of$\ e^{2}\propto t$ was soon ruled out by geochronological and astronomical data \cite{dyson}. Dicke \cite{dicke1957, dicke1961} also provided a simple anthropic explanation for our observation
of some of the large number coincidences, although Dirac seems to have been only partly convinced by them \cite{barrowuni} because he believed that life would always continue once it arose in the universe \cite{farm, barrowuni}.

In more recent times, the main theoretical motivation for studying varying $\alpha $ has come from proposed extensions of the standard model, which often allow one or more of the observed constants of nature to vary in time. For instance, in string theory one generically finds that at low energy the theory contains a scalar field, typically controlling the size of the compact dimensions, which couples directly to matter. In this context, all the observed (3-dimensional) constants of nature will become dynamical quantities sensitive to the size of the internal space \cite{extra}. In fact the existence of time variation in physical constants is probably one of the most robust predictions of such theories (though of course it tells us nothing about the size of the variations to expect). More generally, the beliefs that space has more than three dimensions leads us to expect that the true constants of nature are not the three-dimensional 'shadows' that we observe.

At present, however, the most compelling reason to study varying-$\alpha $ theories is that there has been continuing observational evidence from studies of quasar spectra at high redshift that are consistent with $\alpha $ having changed very slowly over cosmological timescales. The direct experimental investigation of varying $\alpha $ is multi-faceted, and we refer to the recent reviews of the field for full details \cite{uzan11, chiba11, murphy04}. In brief, there are several different probes of variations of $\alpha $, each with their own strengths and limitations. At the present time one can place strong bounds on the variation of $\alpha $ today from precision atomic clock experiments. Although these provide the greatest sensitivity to the local rate of $\alpha $ variation, cosmologically they are of limited value because they only bound $\Delta {}\alpha /\alpha =(\alpha (z)-\alpha _{0})/\alpha _{0}$, where $\alpha _{0}$ is the value today, over a timescale of a few years at most. The 1.8 billion year old natural nuclear reactor at the Oklo uranium mine in Gabon is also extremely sensitive to the value of $\alpha $ at that time \cite{shly} because of the need to preserve a special resonant energy level for neutron capture, but the sensitivity is complicated by the ambient conditions when the reactor operated \cite{oklo}, sensitivity of the key nuclear resonance level to other coupling constants \cite{dys}, and a best fit to the data that is doubled valued for the past value of $\alpha $  -- one of those value ranges includes a small variation around a null variation but the other solution does not \cite{doub}. At far earlier times one can derive weaker bounds on $\Delta {}\alpha /\alpha $ from the physics of the cosmic microwave background \cite{planck} and big bang nucleosynthesis \cite{iocco09}. These give the earliest indirect observational constraints, bounding $\alpha $ at redshifts $z\approx {10^{3}}$ and $z\approx {10^{9}-}10^{10}$ respectively, but for various reasons the constraints are not very tight (approximately $\Delta{}\alpha /\alpha <10^{-2}-10^{-3}$ at best), and they need a theory of $\alpha $ variation in order to link them to data at low redshift and in the lab and an understanding of links to variations of other fundamental parameters of physics at high redshift \cite{cal}.

The most sensitive probes constrain $\Delta {}\alpha /\alpha $ at $z\approx {1-6}$ from observations of spectral lines significantly affected by relativistic effects in absorbing clouds around distant quasars. Evidence for a slow increase in time, $\Delta \alpha /\alpha =-0.57\times 10^{-5},$ from Keck data was found throughout a long programme of observational work by Webb et al \cite{webbprls}. Subsequent claims of a null result, $\Delta \alpha /\alpha =(-0.06\pm 0.06)\times 10^{-5},$ from a different quasar data set were made by Chand et al \cite{chand} but were subsequently shown to suffer from biases in the data analysis method employed which, when corrected, gave $\Delta \alpha /\alpha =(-0.44\pm 0.16)\times 10^{-5}$, consistent with the earlier results of Webb et al, see ref \cite{murph} for details. More recently, more evidence has been found from quasar spectra \cite{webb11, king12} that $\alpha $ has differed from today by $\sim {}5\times 10^{-6}$, but with a complication. Specifically, it was found for $z>1.8$ that $\Delta {}\alpha /\alpha =(-0.74\pm 0.17)\times 10^{-5}$ using data for the Northern sky from the Keck telescope, but $\Delta{}\alpha /\alpha =(0.61\pm 0.20)\times 10^{-5}$ from data for the Southern sky from the VLT telescope, but with some overlapping data to enable detailed cross calibration of the two detectors. Taken at face value this points to $\alpha $ having a large-scale angular dipole of magnitude $\sim0.6\times 10^{-5}$. Recent observations of a single absorber towards the quasar HE 2217-2818 by Molaro et al \cite{molaro} are consistent with this result. Most recently, a new method to probe the spatial constancy of $\alpha $ in our Galaxy using metal lines found in the spectra of white dwarfs by the Hubble Space Telescope has been introduced by Berengut et al \cite{ber}.

Phenomenological models for varying $\alpha $, like those introduced by Gamow, were usually based on assuming $\alpha $ varies as some power law or logarithm of time and simply writing this variation into the usual equations of physics which were derived under the assumption that $\alpha $ is constant. Most observational bounds in the literature (for a review see \cite{dys2}) use this sometimes questionable approach. The first self-consistent theory of varying $\alpha $ is the generalisation of Maxwell's equations due to Bekenstein \cite{bekenstein82}. This was subsequently extended to a cosmological setting and studied in detail in \cite{sandvik02} by Sandvik, Barrow and Magueijo: we shall refer to it as BSBM theory. It provides a self-consistent cosmological theory of varying $\alpha $ in the same way that the Jordan-Brans-Dicke theory does for varying $G$. It has been studied in a range of cosmological and astrophysical situations in refs. \cite{vuc2} and similar ideas were used to create self-consistent theories of varying electron mass in ref \cite{electron} and produce extensions of the Weinberg-Salam theory with varying weak and electromagnetic couplings in refs. \cite{ew}.

In the original BSBM model variations in $\alpha $ occur due to a coupling between the electromagnetic field and a massless scalar field $\phi $ with action
\begin{equation} \label{1.1}
S=\int {d^{4}{x}}\sqrt{-g}\left( \frac{1}{2}R-\frac{1}{2}\omega \partial_{a}\phi {}\partial ^{a}\phi +e^{-2\phi}\mathcal{L}_{em}+\mathcal{L}_{m}\right) ,
\end{equation}
where $\omega $ is a coupling constant, $\mathcal{L}_{em}=-\frac{1}{4}F_{ab}F^{ab}$ is the usual electromagnetic Lagrangian and $\mathcal{L}_{m}$ denotes the Lagrangian for the other matter fields in the theory. There is a variable electric charge and so $\alpha $ is given by
\begin{equation} \label{1.2}
\alpha =\alpha _{0}e^{2\phi },
\end{equation}
where $\alpha_{0}$ is a constant which may be taken as the present value of $\alpha $. Notice that, as $\mathcal{L}_{em}=\frac{1}{2}(E^{2}-B^{2})=0$ for pure radiation, variations in alpha are driven solely by the electromagnetic
energy of non-relativistic matter, parametrised by $\zeta _{m}=\mathcal{L}_{em}/\rho _{m}$ where $\rho _{m}$ is the energy density of non-relativistic matter. The cosmology of this model has been extensively studied when $\zeta_{m}<0$. In this case one has the astronomically attractive picture in which $\alpha $ does not grow in the radiation era, grows logarithmically with time in the dust era and asymptotes to a constant value when the expansion starts to accelerate in the $\Lambda $-dominated era. There have also been some studies of extensions to BSBM by the addition of a non-constant potential \cite{barrow08}, or by allowing the coupling to be a function of $\phi $ \cite{barrow12}. However, so far there has been no study which has allowed for both possibilities. Moreover, the case $\zeta _{m}>0$ has not been investigated much even for the original model. This is primarily
because the approximation method used in the previous studies cannot be extended to this case.

In this paper we aim to study the cosmological dynamics of a generalised BSBM model which allows for both an arbitrary coupling and potential function. We shall perform a dynamical systems analysis of the full, coupled equations in a Friedmann-Robertson-Walker (FRW) background. The only previous study of this form is \cite{farajollahi12}, who studied the case of an exponential potential (in this paper section 6.1). This allows us to derive and extend many of the results of the earlier studies in a unified and more rigorous manner. It will also allow us to understand some cases not dealt with in the earlier analysis, notably the $\zeta _{m}>0$ case.

The outline of this paper is as follows. In section 2 the model we shall study is introduced and discussed, while its cosmology in an FRW background is given in section 3. In Section 4 we reformulate this in terms of a dynamical system using expansion-normalised variables for the case of constant $\omega $ coupling. The next two sections then use this formulation to study various case for constant potential (section 5) and non-constant potentials (section 6) respectively. The phase plane analysis for non-constant coupling is more subtle, so in section 7 we will formulate the theory in a slightly different way to allow both a potential and coupling function to be described as a dynamical system. This is then explored in more detail for the the case of an exponential potential. We draw conclusions in section 8. Appendix A gives more details on how our methods can be extended to closed universes, while appendix B gives some approximate solutions valid when the dynamics become dominated by the scalar field.

In this paper we choose units so that $8\pi{G}=c=\hbar=1$.

\section{The model}
The model we shall study in this paper is defined by the following action
\begin{equation} \label{2.1}
S=\int {d^{4}{x}}\sqrt{-g}\left( \frac{1}{2}R-\frac{1}{2}\omega (\phi)\partial _{a}\phi {}\partial ^{a}\phi -V(\phi )+e^{-2\phi}\mathcal{L}_{em}+\mathcal{L}_{m}\right) ,
\end{equation}
where $\mathcal{L}_{em}=-\frac{1}{4}F_{ab}F^{ab}$, $\mathcal{L}_{m}$ is the Lagrangian of the matter fields, and the coupling function $\omega (\phi )$ and the potential $V(\phi )$ are both arbitrary functions of the scalar field $\phi $ that drives variations in $\alpha $ via eq. \eqref{1.2}; the cosmological constant has been absorbed into the potential, $V$. Note that $%
\phi $ does not directly couple to the matter fields. The model is therefore distinct from chameleon theories, where the scalar field typically couples to all the fields. For the theory to satisfy basic stability requirements we should demand that the scalar field has positive energy and is not a ghost field. This can be done by assuming that $\omega (\phi )\geq0 $ and $V(\phi )\geq 0$ (or, more weakly, that the potential is bounded from below). In this paper we shall always make this assumption unless stated otherwise.

The Einstein equations for this theory are easily found by varying the action with respect to the metric and yield
\begin{equation} \label{2.2}
G_{ab}=T_{ab}^{m}+T_{ab}^{\phi }+e^{-2\phi }T_{ab}^{em},  
\end{equation}
where the energy-momentum tensor for each sector of the theory is defined in the usual way by $T_{i}^{ab}=\frac{2}{\sqrt{-g}}\frac{\delta (\sqrt{-g}\mathcal{L}_{i})}{\delta {g_{ab}}}$. For the scalar field this is
\begin{equation} \label{2.3}
T_{ab}^{\phi }=\omega (\phi )\partial _{a}\phi \partial _{b}\phi +g_{ab}%
\mathcal{L}_{\phi },
\end{equation}
with $\mathcal{L}_{\phi }=-\frac{1}{2}\omega (\phi )\partial _{a}\phi{}\partial ^{a}\phi -V(\phi )$, while $T_{ab}^{m}$ and $T_{ab}^{em}$ take their usual forms. Varying the action with respect to the scalar field gives its equation of motion (where $^{\prime }=d/d\phi )$:
\begin{equation}  \label{2.4}
\Box {\phi }+\frac{\omega ^{\prime }(\phi )}{2\omega (\phi )}\partial_{a}\phi \partial ^{a}\phi -\frac{V^{\prime }(\phi )}{\omega (\phi )}=\frac{2}{\omega (\phi )}e^{-2\phi }\mathcal{L}_{em}. 
\end{equation}
It is this equation which directly governs how $\alpha $ evolves. It is missing from attempts to limit the possibility of varying $\alpha $ by simply writing in a time (or space) dependence into the usual equations of physics. Such attempts ignore the energetics of the $\alpha $ variations and their effects on the curvature of spacetime, which are captured by the field equations, \eqref{2.2}. Finally, varying with respect to the gauge potential gives us the generalised Maxwell equation:
\begin{equation} \label{2.5}
\nabla _{b}(e^{-2\phi }F^{ab})=-\frac{\delta \mathcal{L}_{m}}{\delta {}A_{a}}.
\end{equation}
Some points can be made about this theory. Firstly, this is the most general theory of its kind we could write down with second-order equations of motion. In particular, there is no loss of generality in restricting to an exponential coupling: the case with arbitrary coupling to $\mathcal{L}_{em}$ may be reduced to \eqref{2.1} by a field redefinition. In fact, one can reformulate \eqref{2.1} as a field theory with canonical kinetic terms, but arbitrary coupling terms. This will be demonstrated explicitly in section 7.

Secondly, these equations admit a well-posed initial value formulation, at least for analytic $\omega (\phi )$ and $V(\phi )$ (see theorem 10.1.3 of \cite{wald84}). As a classical theory it is therefore free from pathologies. From a quantum mechanical point of view it corresponds to a theory with non-renormalisable interaction terms. This can be seen explicitly by redefining the field $\phi =\phi (\Phi )$ so that the Lagrangian is canonically normalised:
\begin{equation}  \label{2.6}
\mathcal{L}_{\phi }+e^{-2\phi}\mathcal{L}_{em}=-\frac{1}{2}\partial _{a}\Phi {}\partial ^{a}\Phi -\bar{V}(\Phi )+A(\Phi )\mathcal{L}_{em}. 
\end{equation}
Expanding out the function $A(\Phi )$ perturbatively shows that the terms which mix photons and scalars are of the form $A_{n}\Phi ^{n}\mathcal{L}_{em} $, and so are power-counting non-renormalisable. In the original BSBM theory with constant $\omega $ we would define $\phi =\Phi /\sqrt{\omega }$ to get $A(\Phi )=e^{\frac{-2\Phi }{\sqrt{\omega }}}$, so one should view $1/\sqrt{\omega }$ as the coupling constant for these interaction terms. If $\omega $ is large enough then these terms will be suppressed enough so that
photon-scalar mixing will not be observed in experiments. Notice that due to our choice of units $\omega \sim \mathcal{O}(1)$ corresponds to choosing the fundamental energy scale ($\sqrt{\omega }$) to be near the Planck scale, $\omega <<1$ corresponds to sub-Planckian scales. The coupling vanishes in the limit $\omega \rightarrow \infty $. One can place non-cosmological
bounds on $\omega $ from table-top experiments \cite{bekenstein82} and constraints on the polarization of star light \cite{burrage09}; typically these bounds are at best $\omega\gtrsim 10^{-11}$ (corresponding to an energy scale ${E}\gtrsim 10^{9}GeV$).

It is usual to rewrite the RHS of \eqref{2.4} somewhat differently, by defining for a configuration the dimensionless parameter $\zeta $ by
\begin{equation} \label{2.7}
\zeta =\frac{\mathcal{L}_{em}}{\rho },
\end{equation}
and $\zeta _{m}=\mathcal{L}_{em}/\rho _{m}$ for its cosmological value, where $\rho _{m}$ is the energy density of non-relativistic matter. We do this because, as explained in the introduction, non-relativistic matter is the only source term for the scalar field. Since $\mathcal{L}_{em}=\frac{1}{2}(E^{2}-B^{2})$, and the energy density of the electromagnetic field is $\rho_{em}=\frac{1}{2}(E^{2}+B^{2})$, then clearly $\zeta $ may take values in the
interval
\begin{equation} \label{2.8}
-1\leq {}\zeta {}\leq 1.  
\end{equation}
If $\zeta >0$ then the configuration is dominated by electrostatic energy, while a system with $\zeta <0$ is dominated by its magnetostatic energy. In general $\zeta $ will vary from material to material. However, the cosmological value should be approximately constant, at least over the timescales we consider, and in this paper we will always make this assumption.

The value of $\zeta_{m}$ is not easy to estimate for several reasons. Firstly, since the dominant contribution to $\rho_{m}$ comes from dark matter then if dark matter has any electric or magnetic fields it will dominate $\zeta_{m}$. Normally one would expect $\zeta_{DM}$ to be very small if not zero; almost by definition it does not interact with electromagnetic radiation, so it would seem peculiar if $\zeta_{DM}\neq0$. Despite this, it is worth bearing in mind that one cannot rule out that it makes a significant contribution to $\zeta_{m}$ (for limits on the charge or dipole moments of dark matter see \cite{dermott11, sigurdson04}).

Even estimating $\zeta $ for ordinary baryonic matter is not trivial. Naively, one would expect in an atom that the dominant contribution to $\mathcal{L}_{em}$ would come from the Coulomb binding energy of the nucleon, with all other effects subleading. This can be estimated from the Bethe-Weizs\"{a}cker formula
\begin{equation} \label{2.9}
E_{C}\simeq98.25\alpha{}\frac{Z(Z-1)}{A^{\frac{1}{3}}}\mbox{MeV},
\end{equation}
with $\alpha{}\approx \frac{1}{137}$. This would lead one to expect that $\zeta _{b}\approx 10^{-3}$, with the cosmological value an order of magnitude lower at $\zeta _{m}\approx 10^{-4}$ (unless $\zeta _{DM}\neq 0$). However, this simple argument may overestimate its value. In particular, Bekenstein has argued \cite{bekenstein02} that in this theory a careful analysis shows that the Coulomb contribution cancels, so the leading contribution to $\zeta _{b}$ is actually from the much smaller magnetic dipole of the nucleon. This gives a negative \(\zeta_{m}\) with magnitude \(|\zeta_{m}|\approx10^{-6}\).

One of the interesting consequences of theories like \eqref{2.1} is that they generically predict violations of the weak equivalence principle (WEP) \cite{bekenstein82}. It is easy to see why. A fraction of any particle's mass is electromagnetic in origin and thereby depends on $\alpha $. This means that in a spatial gradient of $\alpha $, which one would expect in a gravitational potential through the Einstein equations, the force on a particle falling in a gravitational potential $h$ will have an additional contribution from $E_{C}=|\zeta |M$ of
\begin{equation} \label{2.10}
F=-M\nabla {h}-\nabla {E_{C}}=-M\nabla {h}-\frac{\partial {E_{C}}}{\partial {\alpha }}\nabla {}\alpha =-M\nabla {h}-|\zeta|M\frac{\nabla {\alpha }}{\alpha }, 
\end{equation}
where we have implicitly assumed $E_{C}$ is proportional to $\alpha $, but this is not crucial for the argument. Clearly then if $\zeta _{1}\neq \zeta_{2}$ for two bodies then they will fall differently in the gravitational field and the WEP will be violated \cite{maj, vuc1, wep}.

Now in the Newtonian limit \eqref{2.2} and \eqref{2.4} reduce respectively to 
\begin{equation} \label{2.11}
\nabla ^{2}h=\frac{1}{2}(1+|\zeta |)\rho,\ {}\ \nabla ^{2}\phi -\frac{m^{2}}{\omega _{0}}\phi =\frac{2\zeta }{\omega_{0}}\rho,
\end{equation}
where $m^{2}=V^{\prime \prime }(0)$ is the scalar field mass, and we have ignored the cosmological constant term in Poisson's equation. For a massless scalar field then the scalar field to this order is given precisely by $\phi =\frac{4\zeta }{\omega _{0}}h$. Using this we can estimate the E\"{o}tv\"{o}s parameter $\eta $ for the accelerations, $a_{1}$ and $a_{2}$, of two freely falling bodies of different composition ('1' and '2') on Earth to be
\begin{equation} \label{2.12}
\eta =\frac{2|a_{1}-a_{2}|}{a_{1}+a_{2}}\simeq \frac{8\zeta _{earth}\times\left\vert \zeta _{1}-\zeta _{2}\right\vert }{\omega_{0}}.
\end{equation}
If we took \(\omega_{0}\sim\mathcal{O}(1)\) then the naive value for \(\zeta_{b}\) one would get from the Coulomb model would give an unacceptable large \(\eta\sim10^{-6}\), in gross conflict with the present limits that \(\eta\lesssim{}\mathcal{O}(10^{-13})\) \cite{wepviolation08}. However, Bekenstein \cite{bekenstein02} has shown, through a detailed study of the full non-linear equations, that in this model any WEP violations are at undetectable small levels: \(\eta\sim\mathcal{O}(10^{-19})\) with \(\omega_{0}\sim\mathcal{O}(1)\). This means this model is not in violation with the weak equivalence principle.

\section{Cosmological equations}
Since we are interested in the cosmology of this model we now specialise our study to the case when the metric takes a Friedmann-Robertson-Walker (FRW) form
\begin{equation} \label{3.1}
ds^{2}=-dt^{2}+a^{2}(t)\left[ \frac{dr^{2}}{1-kr^{2}}+r^{2}(d\theta^{2}+\sin ^{2}\theta {}d\phi ^{2})\right] .
\end{equation}
By such a choice $\alpha $ can only have time dependence, so we cannot directly use these results to explain the apparent spatial dipole in $\alpha$. This will be investigated elsewhere.

For this choice of metric it is easy to see that the scalar field equation of motion takes the form
\begin{equation} \label{3.2}
\ddot{\phi}+3H\dot{\phi}+\frac{\omega ^{\prime }(\phi )}{2\omega (\phi )}\dot{\phi}^{2}+\frac{V^{\prime }(\phi )}{\omega (\phi)}=\frac{-2}{\omega(\phi )}e^{-2\phi }\zeta _{m}\rho _{m}.
\end{equation}
The equivalent Friedmann equation is
\begin{equation} \label{3.3}
\frac{\dot{a}^{2}}{a^{2}}=\frac{1}{3}\left( \rho _{m}(1+|\zeta_{m}|e^{-2\phi })+\rho _{r}e^{-2\phi }+\frac{1}{2}\omega (\phi)\dot{\phi}^{2}+V(\phi )\right) -\frac{k}{a^{2}}, 
\end{equation}
while the acceleration equation becomes
\begin{equation} \label{3.4}
\frac{\ddot{a}}{a}=-\frac{1}{6}\rho _{m}(1+|\zeta _{m}|e^{-2\phi })-\frac{1}{3}\rho _{r}e^{-2\phi }-\frac{1}{3}[\omega (\phi)\dot{\phi}^{2}-V(\phi )].
\end{equation}%
As usual we have assumed that the matter may be modelled as a perfect fluid. The 2nd term multiplying $\rho _{m}$ in these equations arises because, by definition, non-relativistic matter of density $\rho _{m}$ has electromagnetic energy component $|\zeta _{m}|\rho _{m}$, which couples in the Einstein equations to $e^{-2\phi }$. The continuity equation for matter, with the exception of radiation, is unaffected by the scalar field - in particular $\rho _{m}\propto a^{-3}{}$ as in general relativity with no varying $\alpha $. Since radiation couples directly to the scalar field in the action it is easy to see the continuity equation takes the form
\begin{equation} \label{3.5}
\dot{\rho _{r}}+4H\rho _{r}=2\dot{\phi}\rho _{r}.
\end{equation}
This integrates up immediately to give
\begin{equation} \label{3.6}
\rho _{r}e^{-2\phi }\propto \frac{\rho _{r}}{\alpha }\propto {}\frac{1}{a^{4}}.
\end{equation}
This equation has a number of unusual cosmological implication. Statistical mechanics will give $\rho _{r}\propto T_{r}^{4}$ in the usual way, but the evolution of the temperature with scale factor will be $T_{r}\varpropto\alpha ^{1/4}a^{-1}$ and the temperature-redshift relation becomes
\begin{equation} \label{3.7}
T_{r}=T_{r0}(1+z)\left( \frac{\alpha (z)}{\alpha _{0}}\right) ^{1/4},
\end{equation}
which can be tested by detailed constraints on the CMB temperature with redshift, as has also been discussed in ref. \cite{temp}. The relation \eqref{3.7} also means that the combination $T_{r}^{3}/\rho _{m}$, which determines the entropy per baryon in the standard cosmology with constant $\alpha $, is no longer constant as the universe expands. Instead we have
\begin{equation} \label{3.8}
T_{r}^{3}/\rho _{m}\propto \alpha ^{3/4}.
\end{equation}
Hence, any small change in the value of $\alpha $ between the epoch of deuterium synthesis in the early universe and the present will affect deductions of the range of values of the entropy per baryon (and hence the baryon density) that best fit the observed deuterium abundance and effects at CMB last scattering. The evolution given by eq. \eqref{3.7} also changes the calculated value of the time and redshift when the matter and radiation densities are equal, and hence the location of the peak of the matter power
spectrum. These effects were not included in the uses of the Planck data \cite{planck} to constrain possible variations in $\alpha $ because no underlying theory of $\alpha $ variation was used. In addition, we see that the evolution of a neutrino density will not be affected by the fine structure constraint coupling and will evolve as usual, with $\rho _{\nu}\propto T_{\nu }^{4}$, and the ratio of the photon to neutrino temperature will not remain constant but evolve as the quarter power of the fine structure 'constant':
\begin{equation} \label{3.9}
\frac{T_{r}}{T_{\nu }}\varpropto \alpha ^{1/4}.
\end{equation}
We expect, given the existing observational constraints, that the evolution of $\alpha (z)$ will be small but these deviations from the standard picture, which can be computed in detail once a solution for $\phi (t)$ is found from the Friedmann equations, may lead to new constraints on $\alpha $ variation.

We also expect that there will be a powerful constraint on the possible time-evolution of the electromagnetic gauge coupling from any requirement that 'grand unification' occurs at very high energies, $T\sim 10^{15}GeV$. There have already been claims that the requirement of a triple cross-over of the effective interaction strengths of the strong and electroweak couplings was evidence of the need for supersymmetry. However, the addition of an intrinsic time (and hence temperature) evolution over and above that induced
by the quantum vacuum effects would likely destroy the possibility of a grand unification of interaction strengths unless there was considerable fine tuning of the variations. We suspect that they would be constrained to be extremely small over the period of evolution from about $t\sim 10^{-30}s$ to the present.

Equations \eqref{3.2} and \eqref{3.3} are in general too difficult to solve exactly except in highly idealised cases. Most previous studies have proceeded by making some analytical approximations, such as that the scalar terms in \eqref{3.3} can be neglected. A variant on this theme is explored in appendix B. It is the goal of this paper to understand their qualitative behaviour, without any approximation. Before we do so let us note some general features of the cosmology.

Firstly, with zero potential ($V=0$), it is clear from \eqref{3.3} that the influence of the scalar field on cosmological dynamics is to increase the expansion rate: there is no question of the $\phi $ field causing collapse. Similarly, \eqref{3.4} shows that it cannot cause the universe to accelerate, and so cannot be a source of early inflation or late-time accelerated expansion of the universe. Obviously these conclusions may be changed by the addition of a potential.

Secondly, it is important to note that if we do not specify the potential or coupling function then we cannot hope to say much about the cosmological dynamics. In fact, given an observed expansion history for $H(t)$ and $\alpha (t)$ it is always possible to reconstruct functions $V(\phi )$ and $\omega (\phi )$ which lead to this history. This can be seen by noting that equations \eqref{3.2} and \eqref{3.3} can be rewritten as
\begin{align} 
& \frac{1}{2}\omega \dot{\phi}^{2}+V=3H^{2}+\frac{3k}{a^{2}}-\rho_{m}(1+|\zeta _{m}|e^{-2\phi })-\rho _{r}e^{-2\phi }\equiv f(t), \label{3.10}
\\
& \frac{1}{2}\dot{\omega}\dot{\phi}^{2}+\dot{V}=-\dot{\phi}(\ddot{\phi}+3H\dot{\phi})\omega -2\zeta _{m}e^{-2\phi }\rho_{m}\dot{\phi}\equiv g(t)\omega +h(t). \label{3.11}
\end{align}
Differentiating \eqref{3.10} and using \eqref{3.11} gives
\begin{equation} \label{3.12}
\omega (t)=-\frac{1}{3H\dot{\phi}^{2}}(\dot{f}+2\zeta _{m}e^{-2\phi }\rho_{m}\dot{\phi}),
\end{equation}
which gives $\omega (\phi )$ implicitly. Once we have this we can use \eqref{3.10} to find $V(\phi )$. In principle, we could use this as a solution-generating technique to find a desired solution through the choice of 'designer' potentials (in a similar manner to the literature on exact inflationary solutions).

Thirdly, the vacuum solutions of this theory are easy to understand, since in this case the equations \eqref{3.2}-\eqref{3.3} reduce to the usual equations governing inflation with a single scalar field. We can find the exact solution when $V(\phi )=0$ by noting that \eqref{3.2} can be written as $\frac{d}{dt}(\dot{\phi}\omega ^{1/2}a^{3})=0$, which allows us to find the general solution:
\begin{equation} \label{3.13}
a(t)=a_{0}t^{\frac{1}{3}},\ {}\ \int \sqrt{\omega (\phi )}d\phi =\sqrt{\frac{2}{3}}\ln {t}.
\end{equation}
We can also find the exact solution with radiation present without too much difficulty if one works in conformal time. These solutions are the general attractors when $t\rightarrow {0}$, a conclusion explicitly confirmed by the analysis of section 5.

Finally, it is worth noting an important theorem about the behaviour of $\phi (t)$, first given in \cite{barrow02b}: in the absence of a non-constant potential $\phi $ cannot exhibit oscillatory behaviour (as often might appear to be the case from a linearisation of the equations in $\phi $). The proof is immediate from the scalar equation of motion, \eqref{3.2}. At an extrema where $\dot{\phi}=0$ the sign of $\ddot{\phi}$ is fixed uniquely by $\zeta _{m}$, so $\phi $ may only have a maxima (minima) when $\zeta _{m}>0$ ($\zeta _{m}<0$): it cannot have maxima and minima. In particular, $\phi $ cannot have oscillatory behaviour and solutions showing such behaviour (e.g. in ref \cite{DErev}) are spurious, arising from uncontrolled linearisation of \eqref{3.2}. This result can also be extended to certain classes of potentials. For instance, for an exponential potential $V=V_{0}\exp [\beta{}\phi ]$, when $\zeta _{m}>0$ and $\beta >0$ then $\phi $ can only have maxima; when $\zeta _{m}<0$ and $\beta <0$ it can only have minima.

\section{Dynamical systems analysis with constant coupling, $\protect\omega $}
We shall now perform an analysis of the equations \eqref{3.2}-\eqref{3.4} by the methods of dynamical systems \cite{glendinning94}. This is a well known method which has been applied widely in cosmology, for instance see \cite{DErev, wainwright97, barrow86, goliath99}. We will first look at the case when the coupling function \(\omega(\phi)\) is a constant. The case of a general coupling function will be dealt with in section 7.

The first step is to cast equations \eqref{3.2}-\eqref{3.4} into autonomous form. To do so, define the following expansion-normalised variables\footnote{Note our definition of \(x_{2}\) implicitly assumes that \(V\geq0\). If we wanted to allow for a negative cosmological constant we would have to modify these definitions slightly.}
\begin{equation} \label{4.1}
x_{1}=\frac{\sqrt{\omega}\dot{\phi}}{\sqrt{6}H},\ x_{2}=\frac{\sqrt{V}}{\sqrt{3}H},\ x_{3}=\frac{\sqrt{\rho_{m}|\zeta_{m}|}e^{-\phi}}{\sqrt{3}H},\ x_{4}=\frac{\sqrt{\rho_{r}}e^{-\phi}}{\sqrt{3}H},\ x_{5}=\frac{\sqrt{|k|}}{aH}. 
\end{equation}
We will also define
\begin{equation} \label{4.2}
x_{0}=\frac{\sqrt{\rho_{m}}}{\sqrt{3}H},
\end{equation}
although this variable is not independent of the others because the Friedmann equation reduces to a constraint
\begin{equation} \label{4.3}
1=x_{0}^{2}+x_{1}^{2}+x_{2}^{2}+x_{3}^{2}+x_{4}^{2}-\hat{k}x_{5}^{2}, 
\end{equation}
where \(\hat{k}=k/|k|\) is the sign of the curvature. Physically, these variable are the density parameters of each component in the Friedmann equation. Notice that the sign of all these variables, with the exception of \(x_{1}\), is fixed by the Hubble parameter: this means in an expanding universe they are always positive. In principle \(x_{1}\) can take on either sign. In these variables the fine structure 'constant', \(\alpha\), is given by
\begin{equation} \label{4.4}
\alpha=|\zeta_{m}|\left(\frac{x_{0}}{x_{3}}\right)^{2}, 
\end{equation}
and the associated scalar field by
\begin{equation} \label{4.5}
\phi=\ln\left(\frac{x_{0}}{x_{3}}\right)+\phi_{0}, 
\end{equation}
where $\phi _{0}$ is an arbitrary constant. The Hubble parameter is given by
\begin{equation} \label{4.6}
\left(\frac{H}{H_{0}}\right)^{2}=\left(\frac{x_{0,0}}{x_{0}}\right)^{2}e^{-3N}, 
\end{equation}
where \(H_{0}\) and \(x_{0,0}\) is the Hubble parameter and the value of \(x_{0}\) respectively at time \(N=0\), where $N=\ln {}a$. The acceleration equation \eqref{3.4} reduces to
\begin{equation} \label{4.7}
\frac{\dot{H}}{H^{2}}=-\frac{1}{2}(3+3x_{1}^{2}-3x_{2}^{2}+x_{4}^{2}+\hat{k}x_{5}^{2}). 
\end{equation}
If the potential is not constant then we also need to define the following new variables
\begin{equation} \label{4.8}
\lambda=-\frac{V'}{V},\ {}\ \Gamma=\frac{VV''}{V'^{2}}. 
\end{equation}
We can now derive the evolution equations for each variable. This is most conveniently done if one uses the number of e-folds, $N=\ln {}a$, as the time coordinate; it is better behaved than the proper time, $t $, since the interval \(t\in[0,\infty)\) is mapped to \(N\in(-\infty,\infty)\).

Using \eqref{3.2}-\eqref{3.4} and \eqref{4.3}-\eqref{4.7} it is easy to show that the evolution equations for the autonomous variables are
\begin{align}
&\frac{dx_{1}}{dN}=\frac{1}{2}x_{1}(-3+3x_{1}^{2}-3x_{2}^{2}+x_{4}^{2}+\hat{k}x_{5}^{2})-\sqrt{\frac{6}{\omega}}\hat{\zeta}_{m}x_{3}^{2}+\sqrt{\frac{3}{2\omega}}\lambda{}x_{2}^{2}, \label{4.9}
\\
&\frac{dx_{2}}{dN}=-\sqrt{\frac{3}{2\omega}}\lambda{}x_{1}x_{2}+\frac{1}{2}x_{2}(3+3x_{1}^{2}-3x_{2}^{2}+x_{4}^{2}+\hat{k}x_{5}^{2}), \label{4.10}
\\
&\frac{dx_{3}}{dN}=-\sqrt{\frac{6}{\omega}}x_{1}x_{3}+\frac{1}{2}x_{3}(3x_{1}^{2}-3x_{2}^{2}+x_{4}^{2}+\hat{k}x_{5}^{2}), \label{4.11}
\\
&\frac{dx_{4}}{dN}=\frac{1}{2}x_{4}(-1+3x_{1}^{2}-3x_{2}^{2}+x_{4}^{2}+\hat{k}x_{5}^{2}), \label{4.12}
\\
&\frac{dx_{5}}{dN}=\frac{1}{2}x_{5}(1+3x_{1}^{2}-3x_{2}^{2}+x_{4}^{2}+\hat{k}x_{5}^{2}), \label{4.13}
\\
&\frac{d\lambda}{dN}=-\sqrt{\frac{6}{\omega}}\lambda^{2}(\Gamma-1)x_{1}, \label{4.14}
\end{align}
where we write \(\hat{\zeta}_{m}=\zeta_{m}/|\zeta_{m}|\) for the sign of \(\zeta_{m}\). Note that these do indeed form an autonomous system, because in general, as \(\lambda=\lambda(\phi)\), we can solve implicitly for \(\phi=\phi(\lambda)\) which allows us to write \(\Gamma=\Gamma(\phi)=\Gamma(\lambda)\), closing the system. For a constant or exponential potential, \(\lambda=\mbox{constant}\) and the system is defined fully by the variables in \eqref{4.1}; for a more general system one must also include \(\lambda\). Additional perfect fluids in the Friedmann equation can be included without difficulty.

Now that we have the system cast in autonomous form we can determine its behaviour through the qualitative theory of ordinary differential equations. We first of all determine the stationary points of the system, defined by \(dx_{i}/dN=0\). Usually the late and early time attractors of the system will be amongst these points. To determine their stability we linearise the system \(\dot{x}_{i}=f_{i}(x_{j})\) about the stationary point \(x_{0i}\). Explicitly, if we write \(x_{i}=x_{0i}+y_{i}\) then the linearisation is given by
\begin{equation} \label{4.15}
\dot{y}_{i}=A_{ij}y_{j}\ \mbox{ with }\ A_{ij}=\frac{\partial{}f_{i}}{\partial{}x_{j}}\ \bigg|_{x_{i}=x_{0i}}.
\end{equation}
One can then deduce stability through the eigenvalues of \(A_{ij}\). It is a standard result that if the real part of the eigenvalues of \(A\) are entirely negative then the point is stable, while if any are positive it is unstable. If there is a mixture the point is a saddle point, meaning that it is not an attractor at late times but the solutions can come arbitrarily close to the point during its evolution. Provided there are no zero eigenvalues the Hartman-Grobman theorem guarantees that the behaviour near a stationary point is given by the linear approximation. If there is a eigenvalue with zero real part, and no eigenvalue with positive real part, then stability cannot be decided by the linear terms, and one must go at least to 2nd order to decide.

Note that by diagonalising \eqref{4.15}, it is easy to see that the general solution for \(y_{i}\) is given by
\begin{equation} \label{4.16}
y_{i}=\sum_{j}c_{j}e^{\epsilon_{j}t}\chi_{i,j}, 
\end{equation}
where \(\chi_{i,j}\) is the ith component of the jth eigenvector of \(A\) associated to the eigenvalue \(\epsilon_{j}\), and the \(c_{j}\) are constants. This solution gives the leading order correction to the motion near the stationary point.

It is worth pointing out the limitations of these methods. They do not give one much useful information about the solution at intermediate times which is often the case of most interest. It is also worth noting that strictly speaking the above results only hold in a neighbourhood of a stationary point. As well as tending to a stationary point the late time behaviour might be a limit cycle, or a strange attractor (the last case is excluded for two dimensional systems by the Poincar\'e-Bendixson theorem \cite{glendinning94}). If the variables are not compact there may also be stationary points at infinity. For an expanding universe with constant potential, limit cycles are ruled out by the arguments of section 3: for there to be one \(x_{3}\), and thereby \(\phi\), would need to possess both a maximum and a minimum, which is not possible in this case. 

Despite this, equations \eqref{4.9}-\eqref{4.14} are easy enough to simulate numerically, which allows one to check explicitly its late time behaviour. In the next two sections we shall use both methods to understand the cosmology.

\section{Cosmologies with constant potential, $V$, and constant coupling, $\protect\omega $}
Consider first the case where we have a constant potential. This means $\lambda =0$ and so the system is specified fully by \eqref{4.9}-\eqref{4.13}. Although it is not too difficult to do the phase-plane analysis for the entire system, in view of the large number of variables it will be more enlightening to look at special cases in turn.

\subsection{Dynamics with dust}
The simplest case is when the universe contains only dust and the scalar field with the potential zero, that is only \(x_{1}\) and \(x_{3}\) are non-zero. In this case there are 4 stationary points shown in Table 1. The 1st point is the dust-dominated Einstein-de Sitter universe with constant $\alpha $. Near this point the motion is given by
\begin{equation} \label{5.1}
a(t)=a_{0}t^{\frac{2}{3}},\ {}\ \phi(t)=\mbox{constant}. 
\end{equation}
Since it has a zero eigenvalue, its stability cannot be determined by the linear approximation. The 2nd and 3rd points correspond to a universe dominated entirely by the kinetic energy of the scalar field. Dynamically, they behave like universes with a stiff fluid and \(\alpha(t)\) scaling like a power-law,
\begin{equation} \label{5.2}
a(t)=a_{0}t^{\frac{1}{3}},\ {}\ \phi(t)=\phi_{0}{\pm}\sqrt{\frac{2}{3\omega}}\ln{t}. 
\end{equation}
They are always unstable at late times, although by reversing the time it is easy to see they are the attractor solutions at early times. Point 4 again corresponds to a universe dominated by the scalar field which evolves in a power-law fashion as
\begin{equation} \label{5.3}
a(t)=a_{0}t^{\frac{2\omega}{8+3\omega}},\ {}\ \phi(t)=\phi_{0}+\frac{8}{8+3\omega}\ln{t}. 
\end{equation}
Notice that this exponent takes values between \(\frac{1}{3}\) to \(\frac{2}{3}\) so it expands faster than points 2 and 3, but slower than point 1. From the eigenvalues we see that this is a saddle point at late time (notice that since point 4 expands faster than a stiff universe it cannot be the early-time attractor either).

\begin{table} [t]
\begin{center}
  \begin{tabular}{| l || c | c | c | c | c | p{3.5cm} | }
    \hline
   Stationary point (SP) & \(x_{0}\) & \(x_{1}\) & \(x_{3}\) & Existence & Eigenvalues & Stability  \\ \hline
    \(1\) & \(1\) & \(0\) & \(0\) & all \(\omega\), \(\hat{\zeta}_{m}\) & \(0\), \(-\frac{3}{2}\) & transcendentally stable \\ 
    \(2\) & \(0\) & \(1\) & \(0\) & all \(\omega\), \(\hat{\zeta}_{m}\) & \(3\), \(\frac{3}{2}-\sqrt{\frac{6}{\omega}}\) & unstable node \(\omega>\frac{8}{3}\),          saddle point \(\omega<\frac{8}{3}\) \\ 
    \(3\) & \(0\) & \(-1\) & \(0\) & all \(\omega\), \(\hat{\zeta}_{m}\) & \(3\), \(\frac{3}{2}+\sqrt{\frac{6}{\omega}}\) & unstable node \\
    \(4\) & \(0\) & \(\sqrt{\frac{8}{3\omega}}\) & \(\sqrt{\frac{3\omega-8}{3\omega}}\) & \(\omega>\frac{8}{3}\), \(\hat{\zeta}_{m}=-1\) & \(\frac{8}{\omega}\), \(\frac{4}{\omega}-\frac{3}{2}\) & saddle point  \\  \hline
\end{tabular} 
\end{center}
\caption{Stationary points for a universe with dust and a scalar field. The variables are defined in \eqref{4.1} and \eqref{4.2}. Point 1 (the Einstein-de Sitter solution) is the attractor when \(\zeta_{m}<0\), while the late-time behaviour is singular when \(\zeta_{m}>0\).}
\end{table}

To determine the non-linear stability of the dust stationary point we follow the procedure outlined in \cite{barrow86} for the stability analysis with a zero eigenvalue. The first step is to split the system into critical and non-critical variables, where the critical variables are given by the eigenvector of the zero eigenvalue. If we write \(z_{0}\) for the critical variable, and \(z_{i}\) for the non-critical variables (where \(i=1,2,...,n-1\)) then the system \eqref{4.9}-\eqref{4.13} will have been put in the from
\begin{align}
&\dot{z}_{0}=q_{0}(z_{0},...,z_{n-1}), \label{5.4}
\\ 
&\dot{z}_{i}=p_{i}z_{0}+p_{ij}z_{j}+q_{i}(z_{0},...,z_{n-1}), \label{5.5}
\end{align}
where \(q_{0}\) and \(q_{i}\) are of quadratic or higher order in the variables. The system is in canonical form if in addition \(p_{i}=0\). In general one always has the freedom to put the system into this form by an additional (non-linear) transformation (see \cite{barrow86} for details). Once this is done the stability of the point is determined by the leading term of \(q_{0}(z_{0},0,...,0)\). If this term is of the form \(gz_{0}^m\) (\(m\geq2\)) then the stationary point it is unstable if \(m\) is even, or \(m\) is odd and \(g>0\). It is asymptotically stable if \(m\) is odd and \(g<0\), and transcendentally stable if \(q_{0}(z_{0},0,...,0)=0\). By transcendentally stable we mean that the solution approaches a neighbourhood of the stationary point at late times, but \(|\boldmath{z(t)}|\nrightarrow{0}\) as \(t\rightarrow{\infty}\). In fact, at late times these solutions approach \(z_{0}=\mbox{constant}\), \(z_{i}=0\) instead of the stationary point itself. 

In our case these methods are quite easy to apply. One can easily check that \(y_{3}\) is the critical variable, and the perturbation equations about \((0,0)\) are already in the required form:
\begin{equation} \label{5.6}
\dot{y}_{3}=-\sqrt{\frac{6}{\omega}}y_{1}y_{3}+\frac{3}{2}y_{3}y_{1}^{2},\ {}\ {}\  \dot{y}_{1}=-\frac{3}{2}y_{1}+\frac{3}{2}y_{1}^{3}-\sqrt{\frac{6}{\omega}}\hat{\zeta}_{m}y_{3}^{2}. 
\end{equation}
From this we deduce that the 1st stationary point is transcendentally stable. Note this does not prove it is the global attractor of the system: there may be other stationary points at infinity to which the system evolves. If it is the attractor, it tells us that at late times \(x_{1}\rightarrow{0}\), but does not tell us the behaviour of \(x_{3}\). In fact, this depends only on the sign of \(\zeta_{m}\), as can be seen from numerical simulations.

When \(\hat{\zeta}_{m}=-1\), we find that \(x_{3}\rightarrow{0}\), regardless of initial conditions or the value of \(\omega\). In particular, it decays as \(x_{3}\sim{}\frac{1}{\sqrt{N}}\) for large \(N\). This is in agreement with earlier analysis of this model \cite{barrow02}: the universe tends to a dust-dominated universe, with \(\alpha(t)\) growing like \(\ln{}t\). That the point \((0,0)\) is indeed the global attractor of this system can also be seen very clearly from the phase-plane diagram shown in Figure 1.

\begin{figure} [t]
\begin{center}
\includegraphics[scale=1, width=6.5cm, height=3.5cm]{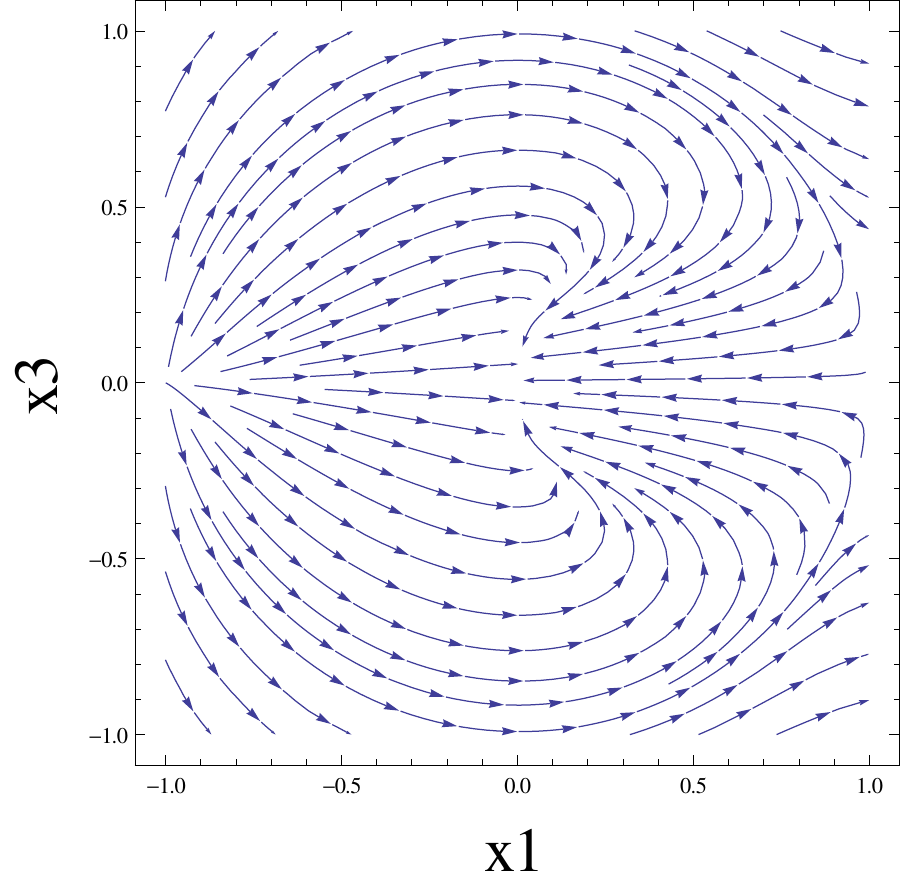}
\end{center}
\caption{Phase plane diagram for the model with dust and a scalar field where \(\hat{\zeta}_{m}=-1\) (\(\omega=\frac{3}{2}\)). The Einstein-de Sitter point \((0,0)\) is the global attractor for all physical values of this system.}
\end{figure}

\begin{figure} [t]
\begin{center}
\includegraphics[scale=1, width=6.5cm, height=3.5cm]{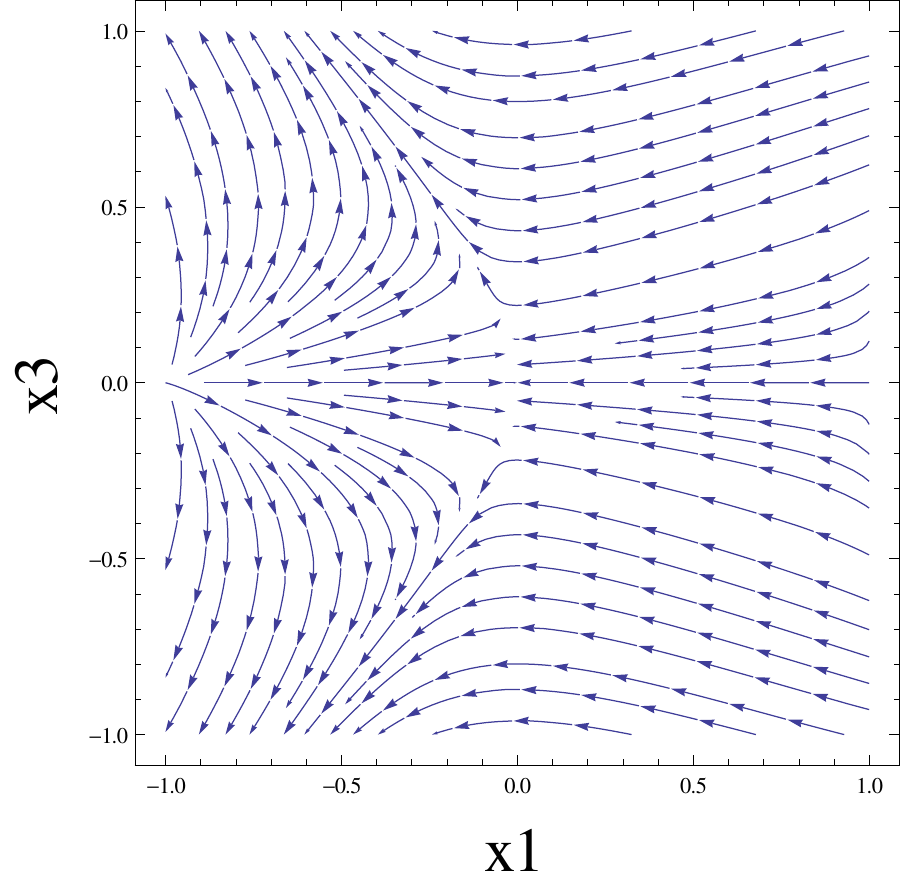}
\end{center}
\caption{Phase plane diagram for the model with dust and a scalar field where \(\hat{\zeta}_{m}=1\) (\(\omega=\frac{3}{2}\)). The Einstein-de Sitter point \((0,0)\) is now a saddle for the system, with the attractors at infinity. Note that physical values of the parameters must lie inside the circle \(x_{1}^{2}+x_{3}^{2}=1\).}.
\end{figure}

The case of \(\hat{\zeta}_{m}=1\) is more complex. Numerically, for reasonable initial data, one finds that initially \(x_{1}\) decreases and \(x_{3}\) grows slowly. At early times, it behaves as a dust-dominated universe with decreasing \(\alpha\). This is only true for low \(N\) though; eventually \(x_{3}\) will be large enough such that the scalar field makes non-negligible contributions to the Friedmann equation and the approximation breaks down. This is why the methods of \cite{barrow02} cannot be used consistently in this case. Following the evolution numerically, one finds that at some critical e-fold \(N\) (proportional to \(1/\omega\)) \(x_{3}\) and \(x_{1}\) rapidly increase and the dynamics become dominated by the scalar field. It is not easy to extract more definite information than this, as the numerical simulations break down at this point. If we look at the phase-plane diagram for this system (Figure 2), we see that the system flows to the circle \(x_{0}=0\). From \eqref{4.6}, this means that at some finite value of the scale factor the Hubble rate diverges, and the universe encounters some type of finite-time singularity.

It is important to note that this somewhat pathological behaviour is not necessarily a practical problem for finding a realistic cosmology with decreasing \(\alpha\). Any non-zero curvature or cosmological constant will cause the scalar field to freeze in once they dominate the dynamics (section 5.2 and 5.3), so the scalar field can only dominate if it has enough time to do so in the matter era \cite{anthr}. In practice, provided that the scalar field does not dominate within \(N\approx{}10\) there will be no problem. This can be satisfied for \(\omega\gtrsim\mathcal{O}(10^{-2})\).

\subsection{Dynamics with a cosmological constant}

\begin{table} [t]
\begin{center}
  \begin{tabular}{| l || c | c | c | c | c | p{3.2cm} | p{3.5cm} | }
    \hline
    SP & \(x_{0}\) & \(x_{1}\)  & \(x_{2}\) & \(x_{3}\) & Existence & Eigenvalues & Stability  \\ \hline
    \(1\) & \(1\) & \(0\) & \(0\) & \(0\) & all \(\omega\), \(\hat{\zeta}_{m}\) & \(0\), \(\pm\)\(\frac{3}{2}\) & unstable saddle \\ 
    \(2\) & \(0\) & \(1\) & \(0\) & \(0\) & all \(\omega\), \(\hat{\zeta}_{m}\) & \(3\), \(\frac{3}{2}-\sqrt{\frac{6}{\omega}}\) & unstable node \(\omega>\frac{8}{3}\),       saddle point \(\omega<\frac{8}{3}\) \\ 
    \(3\) & \(0\) & \(-1\) & \(0\) & \(0\) & all \(\omega\), \(\hat{\zeta}_{m}\) & \(3\), \(\frac{3}{2}+\sqrt{\frac{6}{\omega}}\) & unstable node \\ 
    \(4\) & \(0\) & \(\sqrt{\frac{8}{3\omega}}\) & \(0\) & \(\sqrt{\frac{3\omega-8}{3\omega}}\) & \(\omega>\frac{8}{3}\), \(\hat{\zeta}_{m}=-1\) &  \(\frac{8}{\omega}\), \(\frac{4}{\omega}\pm{}\frac{3}{2}\) &  saddle point \\ 
    \(5\) & \(0\) & \(0\) & \(1\) & \(0\) & all \(\omega\), \(\hat{\zeta}_{m}\) & \(-3\), \(-\frac{3}{2}\) & stable node \\  \hline
\end{tabular}
\end{center}
\caption{Stationary points for a universe with dust, positive cosmological constant and a scalar field. Point 5 (the de Sitter solution) is the global attractor for the system.}
\end{table}

A simple extension is to consider the addition of a positive cosmological constant, \(\Lambda\), into the dynamics. This changes the behaviour radically. There are now 5 types of stationary points shown in Table 2. In addition to the four stationary points found in section 5.1, there is also a new one (point 5) corresponding to the de Sitter universe
\begin{equation} \label{5.7}
a(t)=a_{0}e^{Ht},\ {}\ \phi(t)=\mbox{constant}, 
\end{equation}
where \(H=\sqrt{\Lambda/3}\). Since its eigenvalues are strictly negative this is the global attractor for these solutions, a phenomena which is seen very clearly from numerical simulations of the full system.

This confirms the behaviour found in \cite{barrow02} that \(\phi(t)\) quickly asymptotes to a constant once the universe becomes \(\Lambda\)-dominated. In fact, we can use these results to calculate the leading corrections to the solution about the stationary point as from \eqref{4.16} and \eqref{5.7} we have that
\begin{equation} \label{5.9}
x_{1}=\frac{c_{1}}{a^{3}},\ {}\ x_{2}=1+\frac{c_{2}}{a^{3}},\ {}\ x_{3}=\frac{c_{3}}{a^{3/2}}. 
\end{equation}
These can be explicitly solved to yield
\begin{align}
&\ln{a}-\frac{3c_{2}}{a^{3}}=Ht{}\implies{}a(t)\approx{}e^{Ht}\left(1+\frac{c_{2}}{3}e^{-3Ht}+\mathcal{O}(e^{-6Ht})\right), \label{5.10}
\\
&\phi(t)=\phi_{0}\pm{}c_{1}\sqrt{\frac{2\lambda}{\omega}}\int{}\frac{dt}{a^{3}+c_{2}}\approx{}\phi_{0}\pm\frac{c_{1}}{3}\sqrt{\frac{6}{\omega}}e^{-3Ht}+\mathcal{O}(e^{-6Ht}). \label{5.11}
\end{align}
This means \(\phi(t)\) decays exponentially fast on approach to the de Sitter point. In fact, $\phi $ will asymptote to a
constant whenever the background expansion is dominated by an effective fluid stress with $\rho +3p\leq 0$ which causes the expansion to accelerate \cite{barrow02}.
  
\subsection{Dynamics with curvature}
\begin{table} [t]
\begin{center}
  \begin{tabular}{| l || c | c | c | c | c | c | p{3.5cm} | }
    \hline
   SP & \(x_{0}\) & \(x_{1}\)  & \(x_{3}\) & \(x_{5}\) & Existence & Eigenvalues & Stability  \\ \hline
   \(1\) & \(1\) & \(0\) & \(0\) & \(0\) & all \(\omega\), \(\hat{\zeta}_{m}\) & \(0\), \(-\frac{3}{2}\), \(\frac{1}{2}\) & unstable saddle \\ 
   \(2\) & \(0\) & \(1\) & \(0\) & \(0\) & all \(\omega\), \(\hat{\zeta}_{m}\) & \(3\), \(2\), \(\frac{3}{2}-\sqrt{\frac{6}{\omega}}\) & unstable node \(\omega>\frac{8}{3}\), saddle point \(\omega<\frac{8}{3}\) \\ 
   \(3\) & \(0\) & \(-1\) & \(0\) & \(0\) & all \(\omega\), \(\hat{\zeta}_{m}\) & \(3\), \(2\), \(\frac{3}{2}+\sqrt{\frac{6}{\omega}}\) & unstable node \\ 
   \(4\) & \(0\) & \(\sqrt{\frac{8}{3\omega}}\) & \(\sqrt{\frac{3\omega-8}{3\omega}}\) & \(0\) & \(\omega>\frac{8}{3}\), \(\hat{\zeta}_{m}=-1\) &  \(\frac{8}{\omega}\), \(\frac{4}{\omega}-\frac{3}{2}\), \(\frac{4}{\omega}+\frac{1}{2}\) & saddle point \\ 
   \(5\) & \(0\) & \(0\) & \(0\) & \(1\) & all \(\omega\), \(\hat{\zeta}_{m}\) & \(-2\), \(-1\), \(-\frac{1}{2}\) & stable node \\  \hline
\end{tabular}
\end{center}
\caption{Stationary points for a universe with dust, negative curvature and a scalar field. Point 5 (the Milne solution) is the global attractor for this system.}
\end{table}

We can also consider the effects of adding curvature to the dynamics in a similar manner. For an open universe (\(k<0\)) the stationary points are given in Table 3. The new stationary point is the curvature-dominated Milne universe with solution
\begin{equation} \label{5.12}
a(t)=a_{0}t,\ {}\ \phi(t)=\mbox{constant}, 
\end{equation}
which is the global attractor for this system. As far as the scalar field is concerned, the effects of curvature is very similar to a cosmological constant. The only difference is that the freeze in of \(\phi(t)\) happens a little slower. One can see this by calculating the first order corrections to the motion near the stationary point. One will find that
\begin{align}
&a(t)+c_{5}\ln{}a(t)=\sqrt{|k|}t, \label{5.15}
\\
&\phi(t)=\phi_{0}\pm{}c_{1}\sqrt{\frac{6|k|}{\omega}}\int{}\frac{dt}{a^{2}(a+c_{5})}\approx{}\phi_{0}\pm{}\frac{c_{1}}{|k|}\sqrt{\frac{3}{2\omega}}\frac{1}{t^{2}}+\mathcal{O}\left(\frac{1}{t^{3}}\right), \label{5.16}
\end{align}
where \(c_{1}\) and \(c_{5}\) are constants. In general \(\phi(t)\) decays like a power-law of time in the presence of curvature. 

The case of a closed universe is not so simple, since the variables \eqref{4.1} are no longer compact. Indeed, they diverge at the point of maximum expansion. One can avoid the problem by changing the definition of the variables \eqref{4.1} to avoid this. This is done in appendix A. These results show that closed universes undergo the same collapse as in general relativity, with \(\alpha\) diverging in the collapse as a power-law of time.

\subsection{Dynamics with radiation}

As a final case, let us study the addition of radiation to the dynamics. It is easy to see this does not alter the late-time asymptotes, so for simplicity let us just consider the case of radiation and dust. The stationary points for this case are given in Table 4. For the dust-dominated stationary point, we can use the methods of section 5.1 to show this is also transcendentally stable, as one would expect since radiation is only important at late time. The 5th point corresponds to a radiation-dominated Tolman universe:
\begin{equation} \label{5.17}
a(t)=a_{0}t^{\frac{1}{2}},\ {}\ \phi(t)=\mbox{constant}. 
\end{equation}
This is only a saddle point as one would expect. The last stationary point is rather interesting, in particular it it a physical stationary point provided that \(\omega<8\). It evolves as a radiation-dominated universe with \(\alpha\) growing like a power-law
\begin{equation} \label{5.18}
a(t)=a_{0}t^{\frac{1}{2}},\ {}\ \phi(t)=\phi_{0}+\frac{1}{4}\ln{t}. 
\end{equation}
It is the analogous solution to the radiation-dominated solutions found in \cite{barrow02}, although unlike in that case this is an exact solution of the full set of equations. Like the Tolman solution it is also a saddle point. At early times then, depending on the initial conditions, it is possible for the system to spend much of its time near this point, not the radiation-dominated one.
\begin{table} [t]
\begin{center}
  \begin{tabular}{| l || c | c | c | c | c | c | p{3.2cm} | }
    \hline
    SP & \(x_{0}\) & \(x_{1}\)  & \(x_{3}\) & \(x_{4}\) & Existence & Eigenvalues & Stability  \\ \hline
    \(1\) & \(1\) & \(0\) & \(0\) & \(0\) & all \(\omega\), \(\hat{\zeta}_{m}\) & \(0\), \(-\frac{3}{2}\), \(-\frac{1}{2}\) & transcendentally stable \\ 
    \(2\) &  \(0\) & \(1\) & \(0\) & \(0\) & all \(\omega\), \(\hat{\zeta}_{m}\) & \(3\), \(1\), \(\frac{3}{2}-\sqrt{\frac{6}{\omega}}\) & unstable node \(\omega>\frac{8}{3}\), saddle point \(\omega<\frac{8}{3}\) \\ 
    \(3\) &  \(0\) & \(-1\) & \(0\) & \(0\) & all \(\omega\), \(\hat{\zeta}_{m}\) & \(3\), \(1\), \(\frac{3}{2}+\sqrt{\frac{6}{\omega}}\) & unstable node \\ 
    \(4\) & \(0\) & \(\sqrt{\frac{8}{3\omega}}\) & \(\sqrt{\frac{3\omega-8}{3\omega}}\) & \(0\) & \(\omega>\frac{8}{3}\), \(\hat{\zeta}_{m}=-1\) &  \(\frac{8}{\omega}\), \(\frac{4}{\omega}-\frac{3}{2}\), \(\frac{4}{\omega}-\frac{1}{2}\) & saddle point \\
    \(5\) & \(0\) & \(0\) & \(0\) & \(1\) & all \(\omega\), \(\hat{\zeta}_{m}\) & \(-1\), \(1\), \(\frac{1}{2}\) & saddle point \\ 
    \(6\) & \(0\) & \(\sqrt{\frac{\omega}{24}}\) & \(\sqrt{\frac{\omega}{12}}\) & \(\sqrt{1-\frac{\omega}{8}}\) & \(\hat{\zeta}_{m}=-1\), \(\omega<8\) & \(1\), \(-\frac{1}{2}\pm\frac{\sqrt{2(\omega-6)}}{4}\) & saddle point \\ \hline
\end{tabular}
\end{center}
\caption{Stationary points for a universe with dust, radiation and a scalar field. At early times the system comes near points 5 or 6 depending on the initial conditions. At late times the system becomes dust-dominated.}
\end{table}
\section{Cosmologies with non-constant potential, V($\protect\phi $)$,$ and constant $\protect\omega $}

We now turn to the case when \(V(\phi)\) itself has non-trivial dynamics. Since in general the coupling term provides only small corrections to the Friedmann equation we would expect that the evolution of the scale factor is similar to that in an uncoupled, quintessence model. This turns out to be the case, and this means that, unlike in section 5, the scalar field always modifies the background evolution in a non-trivial manner. 

Given this, we might imagine one could hope to drive variations in \(\alpha\) and a time-varying dark energy with the same scalar field. While in principle this is possible, it does not seem easy to build a phenomenologically viable theory along these lines \cite{barrow08}. It is not difficult to see why. If the universe does accelerate at late times then it must become potential dominated, so the coupling terms in \eqref{3.2} and \eqref{3.3} may be neglected. This means at late times \(\phi\) obeys
\begin{equation} \label{6.1}
\ddot{\phi}+3H\dot{\phi}+\frac{V'(\phi)}{\omega}\approx0. 
\end{equation}
Moreover, for the field \(\phi\) to cause acceleration it should enter the slow-roll regime where \(\dot{\phi}^{2}>>V(\phi)\) and the \(\ddot{\phi}\) term may be neglected in \eqref{6.1} (these can be seen from the acceleration equation \eqref{3.4}). The scalar field will then at late times be given by
\begin{equation} \label{6.2}
\dot{\phi}\approx{-}\frac{V'(\phi)}{3H\omega}\approx{-}\frac{V'(\phi)}{\sqrt{3V(\phi)}\omega}. 
\end{equation}
In particular, unless the potential is actually constant \(\dot{\phi}\neq0\) at late times. This is not in general observationally acceptable because it will lead to a value of \(\dot{\alpha}/\alpha=2\dot{\phi}\) that is too large to be consistent with observational limits (unless the potential is fine-tuned). This heuristic argument is confirmed explicitly by looking at some special cases.

Despite this, it is still interesting to see the different dynamics which occurs when we have more complicated potentials. We will examine the dynamics of the well known exponential and power-law potentials. For simplicity we will just allow for dust in addition to the scalar field; other components could be included without difficulty.

\subsection{Dynamics with exponential potential}
\begin{table} [t]
\begin{center}
  \begin{tabular}{| p{0.3cm} || p{2.6cm} | p{1.4cm} | p{1.8cm} | p{1.8cm} | p{2cm} | p{3cm} | p{2.5cm} | }
    \hline
   SP & \(x_{0}\) & \(x_{1}\)  & \(x_{2}\) & \(x_{3}\) & Existence & Eigenvalues & Stability  \\ \hline
   \(1\) & \(1\) & \(0\) & \(0\) & \(0\) & all \(\omega\), \(\hat{\zeta}_{m}\) & \(0\), \(\pm{}\frac{3}{2}\) & unstable saddle \\ 
   \(2\) &  \(0\) & \(1\) & \(0\) & \(0\) & all \(\omega\), \(\hat{\zeta}_{m}\) & \(3\), \(3+\beta\sqrt{\frac{3}{2\omega}}\), \(\frac{3}{2}-\sqrt{\frac{6}{\omega}}\) & saddle point \(\omega<\frac{8}{3}\) or \(\beta<-\sqrt{6\omega}\) \\ 
   \(3\) & \(0\) & \(-1\) & \(0\) & \(0\) & all \(\omega\), \(\hat{\zeta}_{m}\) & \(3\), \(3-\beta\sqrt{\frac{3}{2\omega}}\), \(\frac{3}{2}+\sqrt{\frac{6}{\omega}}\) & saddle point \(\beta>\sqrt{6\omega}\) \\ 
   \(4\) & \(0\) & \(\sqrt{\frac{8}{3\omega}}\) & \(0\) & \(\sqrt{\frac{3\omega-8}{3\omega}}\) & \(\omega>\frac{8}{3}\), \(\hat{\zeta}_{m}=-1\) &  \(\frac{8}{\omega}\), \(\frac{4}{\omega}-\frac{3}{2}\), \(\frac{4+2\beta}{\omega}+\frac{3}{2}\) & saddle point \\ 
   \(5\) &\(0\) & \(-\frac{\beta}{\sqrt{6\omega}}\) & \(\sqrt{1-\frac{\beta^{2}}{6\omega}}\) & \(0\) & \(|\beta|<\sqrt{6\omega}\) & \(-3+\frac{\beta^{2}}{\omega}\), \(-3+\frac{\beta^{2}}{2\omega}\), \(-\frac{3}{2}+\frac{\beta(2+\beta)}{2\omega}\) & stable node \(-\sqrt{3\omega}<\beta<\sqrt{1+3\omega}-1\), saddle point otherwise \\ 
   \(6\) & \(\sqrt{1-\frac{3\omega}{\beta^{2}}}\) & \(-\frac{1}{\beta}\sqrt{\frac{3\omega}{2}}\) & \(\sqrt{\frac{3\omega}{2\beta^{2}}}\) & \(0\) & \(\omega<\frac{\beta^{2}}{3}\) & \(\frac{3}{\beta}\), \(-\frac{3}{4}\pm\frac{3}{4}\sqrt{\frac{24\omega}{\beta^{2}}-7}\) & stable node \(\beta<-\sqrt{3\omega}\) \\ 
   \(7\) & \(\sqrt{\frac{3\omega(\beta-1)}{2(2+\beta)^2}}\) & \(-\frac{\sqrt{3\omega}}{\sqrt{2}(2+\beta)}\) & \(\sqrt{\frac{3\omega+4(2+\beta)}{2(2+\beta)^{2}}}\) & \(\sqrt{\frac{\beta(2+\beta)-3\omega}{(2+\beta)^{2}}}\) & \(\hat{\zeta}_{m}=-1\), \(\beta>1\) and \(3\omega<\beta(2+\beta)\) & see \eqref{6.8} & stable node \\ 
   \(8\) & \(\sqrt{\frac{4\beta(2+\beta)-3\omega(5+\beta)}{2(2+\beta)^2}}\) & \(-\frac{\sqrt{3\omega}}{\sqrt{2}(2+\beta)}\) & \(\sqrt{\frac{3\omega+4(2+\beta)}{2(2+\beta)^{2}}}\) & \(\sqrt{\frac{3\omega-\beta(2+\beta)}{(2+\beta)^{2}}}\) & see \eqref{6.6} & see \eqref{6.8} & saddle point \\ \hline
\end{tabular}
\end{center}
\caption{Stationary points for a universe with dust and a scalar field with exponential potential. When \(\beta<0\), points 5 and 6 are the global attractors of the system. When \(\beta>0\), depending on the value of \(\omega\) and \(\zeta_{m}\), the late-time behaviour may be point 5, point 7, or a finite-time singularity.}
\end{table}

The first one we study is the case of an exponential potential
\begin{equation} \label{6.3}
V(\phi)=\Lambda{}e^{\beta{}\phi}, 
\end{equation}
where \(\Lambda\) and \(\beta\) are constants. Although the case \(\beta<0\) is the one of most physical interest we will allow \(\beta\) to take either sign. This reduces to a cosmological constant in the limit \(\beta{}\rightarrow{0}\). Note that \(\lambda=-\beta\) is a constant, so the system is specified by \eqref{4.9}-\eqref{4.13} like in section 5. Solving the equations we find that there are 8 types of stationary points, given in Table 5. The first four stationary points are familiar from section 5, though their stability is a little different here. The next two stationary points are the end-states when the evolution becomes potential-dominated. When \(\beta<0\) at least one of these points is an attractor, hence they represent the late-time evolution of the system. Notice that the 5th point reduces to the de Sitter state in the limit \(\beta\rightarrow0\), while the 6th has no analogue. A key difference to the equivalent, de Sitter, attractor with a cosmological constant is that for both of these points \(x_{1}\neq0\), so the scalar field never freezes in. In fact, the solutions in this limit take the form
\begin{equation} \label{6.4}
a(t)=a_{0}t^{\frac{2\omega}{\beta^{2}}},\ {}\ \phi(t)=\phi_{0}-\frac{2}{\beta}\ln{t}, 
\end{equation}
for the first one and
\begin{equation} \label{6.5}
a(t)=a_{0}t^{\frac{2}{3}},\ {}\ \phi(t)=\phi_{0}-\frac{2}{\beta}\ln{t}, 
\end{equation}
for the second. The key point is that this system tends to a universe where both the scale factor and \(\alpha\) grow as a power law of time: \(\alpha=\alpha_{0}t^{-4/\beta}\). Such fast evolution of \(\alpha(t)\) can be used to place strong bounds on \(\beta\), as was first done in \cite{barrow08}.

The last two stationary points are rather complex. Point 7 exists when \(\hat{\zeta}_{m}=-1\), \(\beta>1\) and \(3\omega<\beta(2+\beta)\) (the last two conditions come from demanding that the variables are real). Point 8 exists when \(\hat{\zeta}_{m}=1\) and
\begin{equation} \label{6.6}
3\omega>\beta(2+\beta),\ {}\ 3\omega(5+\beta)<4\beta(2+\beta)\ \mbox{ and } 3\omega+4(2+\beta)>0.  
\end{equation}
Consideration of these conditions shows that it can only exist when \(\beta>0\) or \(\beta<-5/2\). In both cases these are power-law solutions with
\begin{equation} \label{6.7}
a(t)=a_{0}t^{\frac{2(2+\beta)}{3\beta}},\ {}\ \phi(t)=\phi_{0}-\frac{2}{\beta}\ln{t}. 
\end{equation}
The eigenvalues are 
\begin{equation} \label{6.8}
-\frac{6}{2+\beta},\ {}\ -\frac{3(8+6\beta+\beta^{2})}{4(2+\beta)^{2}}\pm\frac{1}{4}\sqrt{\frac{3(24\omega(10+3\omega)+8\beta(9\omega-16)-\beta^{2}(128+21\omega)-32\beta^{3})}{\omega(2+\beta)^{2}}}. 
\end{equation}
We can show that one of these is always a real positive number, or a complex number with positive real part, when \(\beta<0\), or when \(\beta>0\) and \(3\omega>\beta(2+\beta)\). When \(\beta>0\) and \(3\omega<\beta(2+\beta)\) all the eigenvalues are negative or have real negative part. We therefore conclude that point 8 is a saddle point, while point 7 is a stable node. This is confirmed by numerical simulations (Figure 3).

\begin{figure}[htp]
  \begin{center}
    \subfigure[\(\omega=6\)]{\label{fig:edge-a}\includegraphics[scale=0.65]{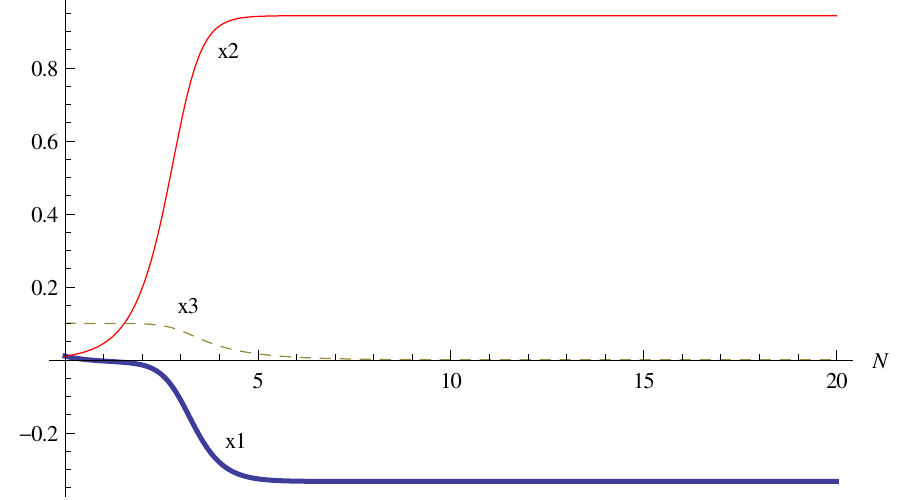}}
    \subfigure[\(\omega=2\)]{\label{fig:edge-b}\includegraphics[scale=0.65]{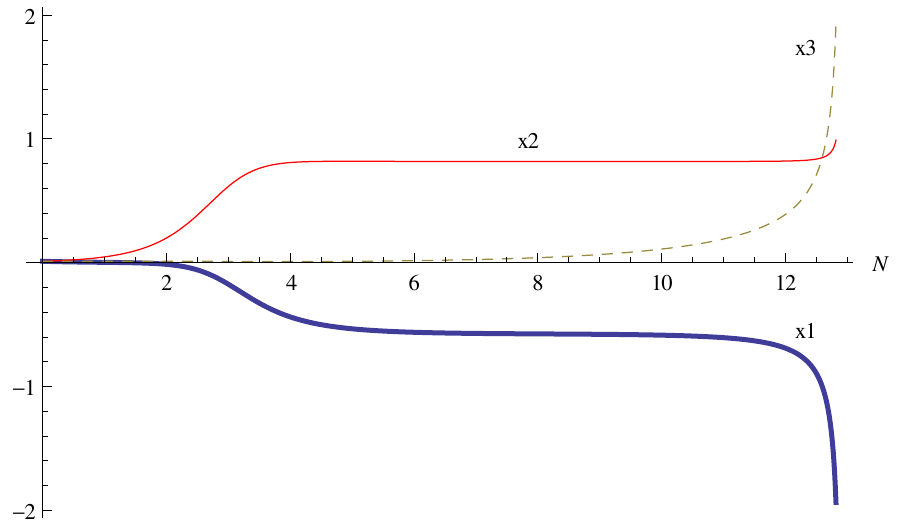}}
  \end{center}
  \caption{Simulations for a universe with dust and a scalar field with exponential potential, \(\beta=2\) and \(\zeta_{m}>0\). The thick blue (lower) curve is the value of \(x_{1}\), the red (upper) of \(x_{2}\) and the dashed yellow (intermediate) of \(x_{3}\). For (a) we have set \(x_{1}=0.01\) and \(x_{3}=0.1\) as our initial conditions, and \(\omega=6\). The solution clearly tends to the attractor point 5. For (b) we set \(x_{1}=x_{3}=0.01\) initially and choose \(\omega=2\). This solution tends to a singularity at a finite value of the scale factor.}
  \label{fig:edge}
\end{figure}

To summarise, when \(\beta<0\) then either point 5 or 6 is the attractor for all \(\omega\), and the late-time behaviour is given by the power-law solutions \eqref{6.4} and \eqref{6.5}. For the less physically interesting case of \(\beta>0\), if \(\zeta_{m}<0\) then point 5 is the attractor provided \(3\omega>\beta(2+\beta)\), otherwise point 7 will be the attractor. By contrast, when \(\zeta_{m}>0\) point 5 is still the attractor if \(3\omega>\beta(2+\beta)\), but if this is not satisfied the solution develops a finite-time singularity. This has been checked using numerical simulations of the equations.

\subsection{Dynamics with power-law potential}
The next potential we examine is when \(V(\phi)\) takes a power-law form
\begin{equation} \label{6.9}
V(\phi)=\Lambda\phi^{n}, 
\end{equation}
where \(\Lambda\) and \(n\) are constants. For this potential \(\lambda\) is no longer a constant, instead \(\lambda=-\frac{n}{\phi}\) and \(\Gamma=1-\frac{1}{n}\). The system is defined by the equations \eqref{4.9}-\eqref{4.14}. Now eq. \eqref{4.14},
\begin{equation} \label{6.10}
\frac{d\lambda}{dN}=\sqrt{\frac{6}{\omega}}\frac{\lambda^{2}}{n}x_{1}, 
\end{equation}
implies that any stationary point must have \(\lambda=0\) or \(x_{1}=0\). The 1st choice gives the same solutions as in Table 2, while the 2nd only has the trivial solution \(x_{1,2,3}=0\). This means the model has the same stationary points as in Table 2 with \(\lambda=0\), plus the point \(x_{0}=1\), \(x_{1,2,3}=0\) and \(\lambda=\mbox{constant}\). The eigenvalues for these points are the same as given in Table 2, but with an extra zero eigenvalue added to all of them (the new point has a double zero eigenvalue and a pair with values \(\pm\frac{3}{2}\), so cannot be stable).

To determine if the de Sitter point -- \(x_{2}=1\), \(x_{0,1,3}=0\), \(\lambda=0\) -- is an attractor we must follow the procedure outlined in section 5.1. We find that the point is (asymptotically) stable when \(n<0\), and unstable for \(n>0\). This means the system is attracted to the de Sitter solution at late times when \(n<0\). When \(n>0\) one finds, from numerical simulations, that a finite-time singularity develops for the system.

That the late-time behaviour is essentially identical to the case with constant potential is not surprising. It is well known that power-law potentials exhibit tracking behaviour: the scalar field tracks the energy density of dust at intermediate times, before dominating entirely at late times. As this is an effect at intermediate redshift it cannot be seen by a phase-plane analysis of the system. For our case, numerical evolution shows, for \(n<0\), that the \(\lambda\) rolls to zero very slowly, and at intermediate redshift the solution is to a good approximation give by \(\lambda\approx{}\mbox{constant}\). From section 6.1, we see that $x_{1}$ and $x_{2}$ are approximately constant over a redshift range where this is valid. Solving in this limit gives
\begin{equation} \label{6.11}
a(t)=a_{0}\exp [At^{2/(2-n)}],\ {}\ \phi (t)=Bt^{2/(2-n)},
\end{equation}
where \(A\) and \(B\) are constants depending on \(\lambda\). Ultimately, in this model one would expect $\alpha (t)$ to have a very fast time variation, in conflict with observations.

\section{Dynamical systems analysis with arbitrary coupling}
We now wish to examine the cosmology when the coupling is a function of the scalar field, first considered with $V=0$ in \cite{barrow12}. For this analysis the formulation of the theory given in section 2 is not optimal. Instead, it is better to make a field redefinition \(\phi=\phi(\Phi)\) so that the action is canonically normalised
\begin{equation} \label{7.1}
S=\int{d^{4}x}\sqrt{-g}\left(\frac{1}{2}R-\frac{1}{2}\partial_{a}\Phi{}\partial^{a}\Phi-\bar{V}(\Phi)+A(\Phi)\mathcal{L}_{em}+\mathcal{L}_{m}\right). 
\end{equation}
Explicitly, this can be done if we make the choice
\begin{equation} \label{7.2}
\int{}\sqrt{\omega(\phi)}d\phi=\Phi, 
\end{equation}
which gives \(\phi(\Phi)\) implicitly. For instance, for the case of a exponential coupling \(\omega=\omega_{0}e^{\mu\phi}\) \cite{barrow12} solving this shows that we should choose \(\Phi\) so that
\begin{equation} \label{7.3}
\phi=\frac{2}{\mu}\ln\left(\frac{\mu{}\Phi}{2\sqrt{\omega_{0}}}\right). 
\end{equation}
It is easy to check that the coupling term \(A(\Phi)\) takes a power-law form, \(A(\Phi)=A_{0}\Phi^{-\frac{4}{\mu}}\), in this case. Similarly, for a power-law coupling function \(\omega=\omega_{0}\phi^{n}\) the correct choice is \(\phi=A'_{0}\Phi^{\frac{2}{2+n}}\), which leads to \(A(\Phi)=e^{-2A'_{0}\Phi^{\frac{2}{2+n}}}\). It is also easy to invert the transformation and go back to the original theory: one simply solves \(A(\Phi)=e^{-2\phi}\) to get \(\Phi=\Phi(\phi)\).

Since these two formulations of the theory are entirely equivalent we are free to study either. We will use this formulation to study the cosmology with arbitrary coupling. In this formulation \(\alpha\) is given by 
\begin{equation} \label{7.4}
\alpha=\frac{1}{A(\Phi)}.
\end{equation}
The equivalent cosmological equations to \eqref{3.2}-\eqref{3.4} are
\begin{align}
&\ddot{\Phi}+3H\dot{\Phi}+\bar{V}'(\Phi)=A'(\Phi)\zeta_{m}\rho_{m}, \label{7.5}
\\
&H^{2}=\frac{1}{3}\left(\rho_{m}(1+|\zeta_{m}|A(\Phi))+\rho_{r}A(\Phi)+\frac{1}{2}\dot{\Phi}^{2}+\bar{V}(\Phi)\right)-\frac{k}{a^{2}}, \label{7.6}
\\
&\dot{H}=-\frac{1}{2}\rho_{m}(1+|\zeta_{m}|A(\Phi))-\frac{2}{3}\rho_{r}A(\Phi)-\frac{1}{2}\dot{\Phi}^{2}+\frac{k}{a^{2}}. \label{7.7}
\end{align}
To cast this into autonomous form we follow the same steps as in section 4. We define autonomous variables by
\begin{equation} \label{7.8}
x_{1}=\frac{\dot{\Phi}}{\sqrt{6}H},\ x_{2}=\frac{\sqrt{\bar{V}}}{\sqrt{3}H},\ x_{3}=\frac{\sqrt{\rho_{m}|\zeta_{m}|A}}{\sqrt{3}H},\ x_{4}=\frac{\sqrt{\rho_{r}A}}{\sqrt{3}H},\ x_{5}=\frac{\sqrt{|k|}}{aH}, 
\end{equation}
and also define \(x_{0}=\sqrt{\rho_{m}/3H^{2}}\). Note that in these variables \(\alpha\) and \(H/H_{0}\) continue to be given by \eqref{4.4} and \eqref{4.6} respectively. The scalar field is gotten by solving \(\alpha(N)=A(\Phi)^{-1}\). If either the potential or the coupling is non-constant we also need to define
\begin{equation} \label{7.9}
\lambda_{V}=-\frac{\bar{V}'}{\bar{V}},\ {}\ \Gamma_{V}=\frac{\bar{V}\bar{V}''}{\bar{V}'^{2}},\ {}\ \lambda_{A}=-\frac{A'}{A},\ {}\ \Gamma_{A}=\frac{AA''}{A'^{2}}. 
\end{equation}
The full evolution equations for this system are
\begin{align}
&\frac{dx_{1}}{dN}=\frac{1}{2}x_{1}(-3+3x_{1}^{2}-3x_{2}^{2}+x_{4}^{2}+\hat{k}x_{5}^{2})-\sqrt{\frac{3}{2}}\lambda_{A}\hat{\zeta}_{m}x_{3}^{2}+\sqrt{\frac{3}{2}}\lambda_{V}x_{2}^{2}, \label{7.10}
\\
&\frac{dx_{2}}{dN}=-\sqrt{\frac{3}{2}}\lambda_{V}x_{1}x_{2}+\frac{1}{2}x_{2}(3+3x_{1}^{2}-3x_{2}^{2}+x_{4}^{2}+\hat{k}x_{5}^{2}), \label{7.11}
\\
&\frac{dx_{3}}{dN}=-\sqrt{\frac{3}{2}}\lambda_{A}x_{1}x_{3}+\frac{1}{2}x_{3}(3x_{1}^{2}-3x_{2}^{2}+x_{4}^{2}+\hat{k}x_{5}^{2}), \label{7.12}
\\
&\frac{dx_{4}}{dN}=\frac{1}{2}x_{4}(-1+3x_{1}^{2}-3x_{2}^{2}+x_{4}^{2}+\hat{k}x_{5}^{2}), \label{7.13}
\\
&\frac{dx_{5}}{dN}=\frac{1}{2}x_{5}(1+3x_{1}^{2}-3x_{2}^{2}+x_{4}^{2}+\hat{k}x_{5}^{2}), \label{7.14}
\\
&\frac{d\lambda_{V}}{dN}=-\sqrt{6}\lambda_{V}^{2}(\Gamma_{V}-1)x_{1}, \label{7.15}
\\
&\frac{d\lambda_{A}}{dN}=-\sqrt{6}\lambda_{A}^{2}(\Gamma_{A}-1)x_{1}. \label{7.16}
\end{align}
In addition, we also have the constraint equation
\begin{equation} \label{7.17}
1=x_{0}^{2}+x_{1}^{2}+x_{2}^{2}+x_{3}^{2}+x_{4}^{2}-\hat{k}x_{5}^{2}. 
\end{equation}
This formalism is considerably more general than the one developed in section 4, and can be used to study the dynamics for any potential or coupling function (although for the case of constant coupling the formalism developed in section 4 is more useful). Before we look at some specific cases, it is worth noting that there are some general stationary points which exist regardless of the detailed form of the potential or coupling. These include the points 
\begin{align}
(x_{0}, x_{1}, x_{2}, x_{3}, x_{4}, x_{5})&=(1,0,0,0,0,0)\ {}\  \mbox{[Einstein-de Sitter point]}, \label{7.18}
\\
&=(0,0,1,0,0,0) \mbox{ and }\lambda_{V}=0\ {}\  \mbox{[de Sitter point]}, \label{7.19}
\\
&=(0,0,0,0,0,1) \mbox{ and }\hat{k}=-1\ {}\  \mbox{[Milne point]}, \label{7.20}
\\
&=(0,0,0,0,1,0)\ {}\  \mbox{[Tolman point]}, \label{7.21}
\\
&=(0,\pm1,0,0,0,0) \mbox{ and }\lambda_{V}, \lambda_{A}=0 \mbox{ or }\Gamma_{V}, \Gamma_{A}=1 \ {}\ 
\\
&\mbox{[scalar dominated point]}. \label{7.22}
\end{align}
In general the stability of these points will depend on the form of the potentials, but in some cases the linear approximation is enough to decide. For instance, the eigenvalues of the last point always include \(3\), so it can never be stable. By contrast, the eigenvalues of the de Sitter point are \(-3\), \(-3/2\), \(-2\), \(-1\) and a double zero eigenvalue. If there is no non-constant potential or coupling then the point is stable; if either potential or coupling is non-trivial its stability must be determined by the non-linear analysis discussed in section 5.1.

\subsection{Case of exponential coupling}
As an application of this formalism let us study the case when the original coupling \(\omega(\phi)\) takes an exponential form: 
\begin{equation} \label{7.23}
\omega(\phi)=\omega_{0}e^{\mu\phi}. 
\end{equation}
As explained above this is equivalent to a theory with a power-law form for \(A(\Phi)=A_{0}\Phi^{-\frac{4}{\mu}}\). Notice this means that \(\lambda_{A}=\frac{4}{\mu\Phi}\) and \(\Gamma_{A}=1+\frac{\mu}{4}\) is a constant. For simplicity we will restrict to the case of a constant potential. As in the previous section the
dynamics is strongly dependent on whether $V$ vanishes or not.

We will first examine the case of a positive cosmological constant. To find the stationary points, note that \eqref{7.16} implies that either \(x_{1}=0\) or \(\lambda_{A}=0\). We can then solve the remaining equations to find that there are only 4 types of stationary points, given in Table 6. For the last point the value of \(\lambda_{A}\) is not fixed.

\begin{table} [t]
\begin{center}
  \begin{tabular}{| l || c | c | c | c | c | c | c | p{3.5cm} | }
    \hline
    SP & \(x_{0}\) & \(x_{1}\) & \(x_{2}\) & \(x_{3}\) & \(\lambda_{A}\) & Existence & Eigenvalues & Stability  \\ \hline
    \(1\) & \(1\) & \(0\) & \(0\) & \(0\) & \(0\) & all \(\omega\), \(\hat{\zeta}_{m}\) & \(0\), \(\pm\frac{3}{2}\) & saddle point \\ 
    \(2\) & \(0\) & \(1\) & \(0\) & \(0\) & \(0\) & all \(\omega\), \(\hat{\zeta}_{m}\) & \(0\), \(3\), \(\frac{3}{2}\) & unstable node \\ 
    \(3\) & \(0\) & \(-1\) & \(0\) & \(0\) & \(0\) & all \(\omega\), \(\hat{\zeta}_{m}\) & \(0\) \(3\), \(\frac{3}{2}\) & unstable node \\ 
    \(4\) & \(0\) & \(0\) & \(1\) & \(0\) & constant & all \(\omega\), \(\hat{\zeta}_{m}\) & \(0\), \(-3\), \(-\frac{3}{2}\) & transcendental stable \\ \hline
\end{tabular} 
\end{center}
\caption{Stationary points for a universe with dust, positive cosmological constant, and a scalar field with exponential coupling. Point 4 (the de Sitter solution) is the attractor for this system.}
\end{table}

\begin{figure}[htp]
  \begin{center}
    \subfigure[\(\mu=0.01\)]{\label{fig:edge-a}\includegraphics[scale=0.65]{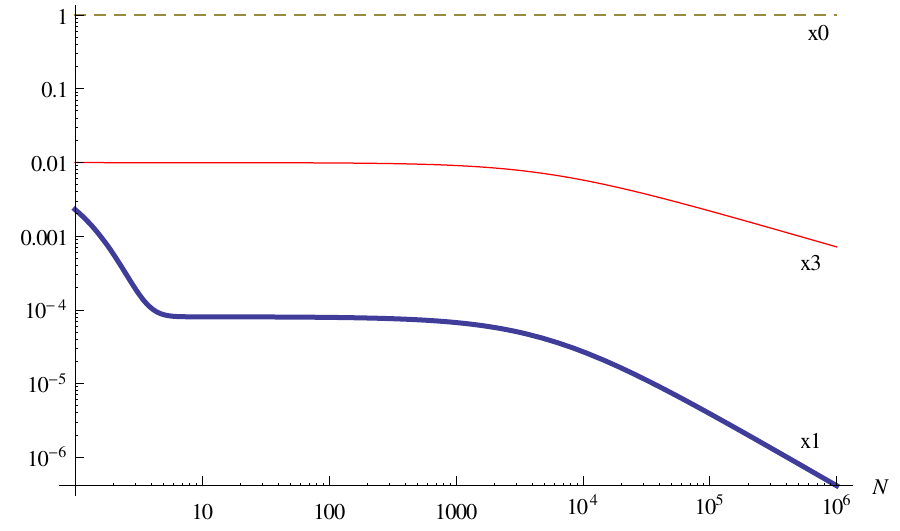}}
    \subfigure[\(\mu=10.01\)]{\label{fig:edge-b}\includegraphics[scale=0.65]{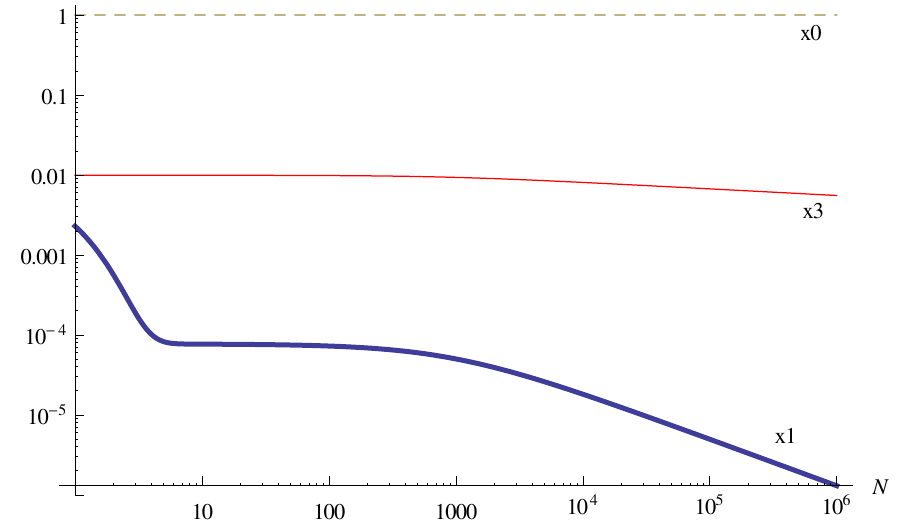}}
  \end{center}
  \caption{Simulations for a universe with dust and a scalar field with exponential coupling, \(\omega_{0}=4\) and \(\zeta_{m}<0\). The thick blue (lower) curve is the value of \(x_{1}\), the red (intermediate) of \(x_{3}\) and the dashed yellow (upper) of \(x_{0}\) in both cases. For both simulations we choose initial conditions \(x_{1}=x_{3}=0.01\). For (a) we have set \(\mu=0.01\), in (b) \(\mu=10.01\). Both show that at late times the system tends to a dust-dominated universe, with \(x_{3}\) rolling slowly to zero and \(x_{1}\) subdominant. The only difference is that (b) shows slower decrease of \(x_{3}\), which corresponds to \(\alpha\) growing more slowly in time for this solution.}
  \label{fig:edge}
\end{figure}
Clearly, from the eigenvalues, the only potentially stable point is the 3rd: the de Sitter point. It has a zero eigenvalue due to the de Sitter solution actually being a curve of critical points given by \(x_{1}=x_{3}=0\) and \(x_{2}=1\) in phase space. Using the methods outlined in section 5 we can show it is transcendentally stable for any value of \(\lambda_{A}\). This is confirmed from numerical simulations, which show that for reasonable initial conditions the solution quickly asymptotes to a de Sitter state, with the value of \(\lambda_{A}\) frozen in close to its initial value. More precisely, if \(\mu>0\) then \(\lambda_{A}\) decreases, while if \(\mu<0\) it increases before freezing in (this can be seen directly from \eqref{7.16}).

The behaviour with dust alone is more complex: in this case the last stationary point vanishes and the only possible attractor is the Einstein-de Sitter point. Since this point has a double zero eigenvalue the techniques of Lyapunov and Malkin described in \cite{barrow86} cannot be applied. Instead we deduce the behaviour from numerical simulations. Before we describe the results, it is worth noting that when one chooses the initial data for the simulations one is not free to specify the initial value of \(\lambda_{A}\) freely once \(x_{1}\) and \(x_{3}\) are chosen. Instead, it is given by (writing in terms of the original parameters)
\begin{equation} \label{7.24}
\lambda_{A}=\frac{2}{\sqrt{\omega_{0}}}\left(\frac{x_{3}}{x_{0}\sqrt{|\zeta_{m}}|}\right)^{\mu/2}. 
\end{equation}
This is in addition to the constraint equation \eqref{7.17}, which in this case enforces \(x_{1}^{2}+x_{3}^{2}\leq1\).

The results we find are as follows. When \(\hat{\zeta}_{m}=-1\) then in all cases \(x_{3}\) and \(x_{1}\) decreases (at least initially), which means that \(\alpha(t)\) increases. When \(\mu>0\) then \(\Phi\) increases and the behaviour is very similar to the \(\omega=\mbox{constant}\) case discussed in section 5.1: the solution tends to an Einstein-de Sitter universe at late times, with \(\alpha\) growing like the logarithm of time (Figure 4). The only difference is that the actual rate of growth depends on \(\mu\); in particular, for larger \(\mu\) the growth is slower. This is in line with the results of \cite{barrow12}.

When \(\hat{\zeta}_{m}=-1\), but \(\mu<0\), then we now have that \(\Phi\) decreases. This is not inconsistent with \(\alpha\) continuing to increase, since as \(\alpha=A^{-1}=\alpha_{0}\Phi^{4/\mu}\), then when \(\mu>0\) an increase of \(\Phi\) leads also to an increase in \(\alpha\), while when \(\mu<0\) the opposite is true. As noted in \cite{barrow12} there is a change in behaviour when \(\mu<-2\), since for this value the simulations develop a singularity in finite time (Figure 6a). This is qualitatively different from the behaviour found in section 5.1, since \(x_{3}\) continues to decrease upon approach to this point. It may be that the solution becomes dominated by the kinetic energy of the scalar field, although the simulations do not have sufficient resolution to show this.

\begin{figure}[htp]
  \begin{center}
    \subfigure[simulation]{\label{fig:edge-a}\includegraphics[scale=0.65]{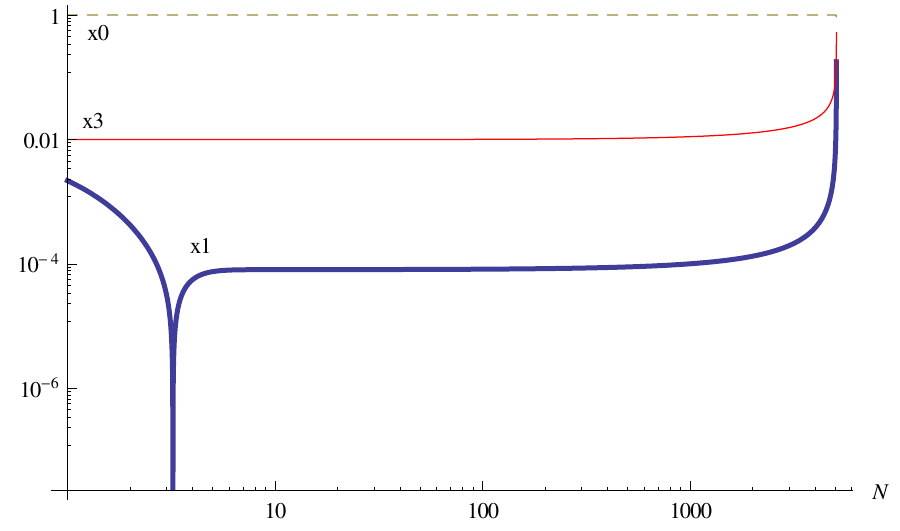}}
    \subfigure[\(\alpha\)]{\label{fig:edge-b}\includegraphics[scale=0.65]{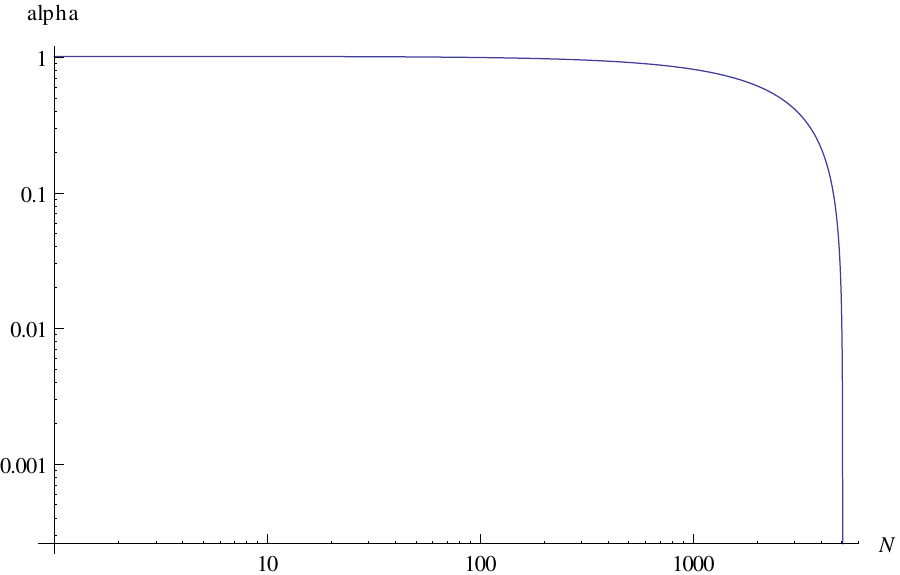}}
  \end{center}
  \caption{Simulation for a universe with dust and a scalar field with exponential coupling, \(\omega_{0}=4\), \(\mu=0.01\) and \(\zeta_{m}>0\). In (a) The thick blue (lower) curve is the magnitude of \(x_{1}\), the red (intermediate) of \(x_{3}\) and the dashed yellow (upper) of \(x_{0}\). We choose initial conditions \(x_{1}=x_{3}=0.01\). The solution early on is like a dust-dominated universe with decreasing \(\alpha\). However, at some critical redshift (\(N=5078\) here) both \(x_{1}\) and \(x_{3}\) rapidly increase, while \(x_{0}\) decreases, and the solution develops a finite-time singularity due to the scalar field. Note that \(x_{1}\) goes through zero and becomes negative, and \(\alpha\) decreases to zero as shown in (b).}
  \label{fig:edge}
\end{figure}

\begin{figure}[htp]
  \begin{center}
    \subfigure[\(\zeta_{m}<0\)]{\label{fig:edge-a}\includegraphics[scale=0.65]{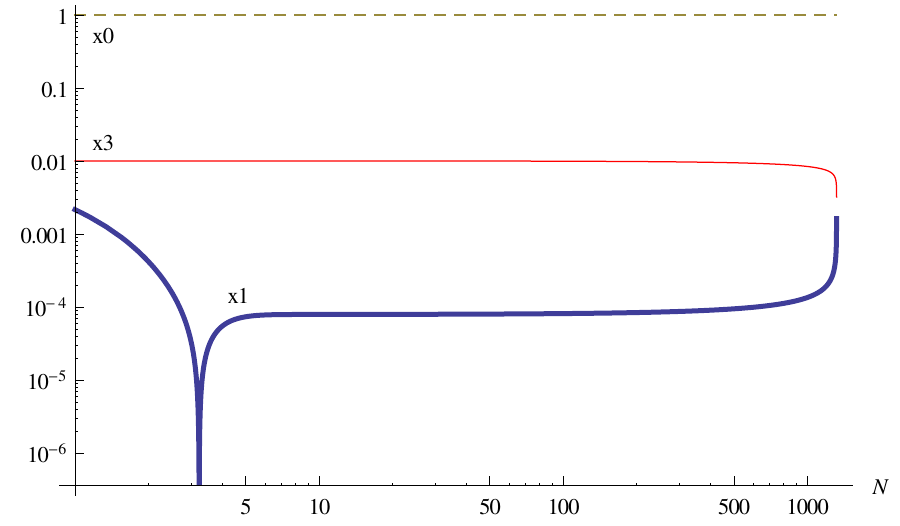}}
    \subfigure[\(\zeta_{m}>0\)]{\label{fig:edge-b}\includegraphics[scale=0.65]{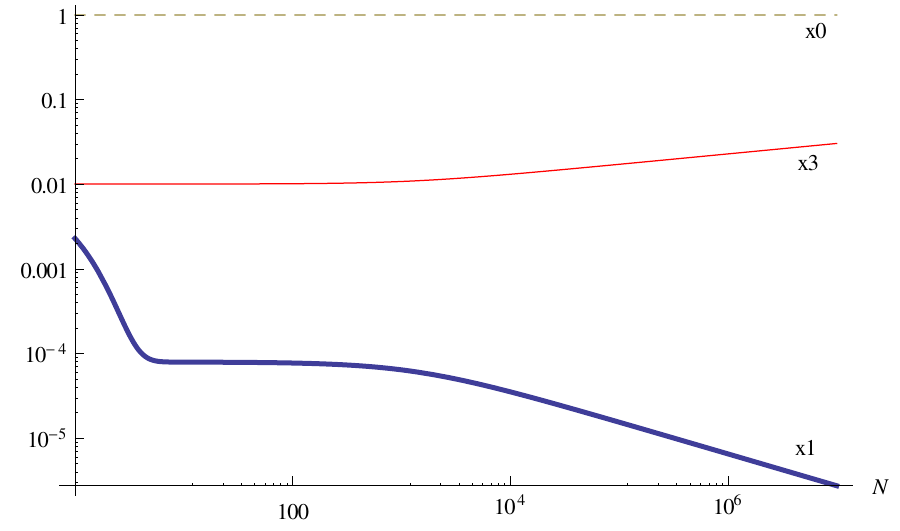}}
  \end{center}
  \caption{Simulations for a universe with dust and a scalar field with exponential coupling, \(\omega_{0}=4\) and \(\mu=-10.1\). The thick blue (lower) curve is the magnitude of \(x_{1}\), the red (intermediate) of \(x_{3}\) and the dashed yellow (upper) of \(x_{0}\). For both simulations we choose initial conditions \(x_{1}=x_{3}=0.01\). For (a), with \(\zeta_{m}<0\), we find that the solution develops a finite-time singularity, but it is qualitatively different from the one found in the \(\zeta_{m}>0\), \(\mu>-2\) case. The only divergence occurs due to the \(x_{1}\) term (which becomes negative), with \(x_{3}\) continuing to decrease. For (b), with \(\zeta_{m}>0\), the future evolution appears non-singular, but \(x_{3}\) continues to increase (in line with \(\alpha\) decreasing) and \(x_{1}\) is always negligible. The late-time solution is one dominated by the scalar field.}
  \label{fig:edge}
\end{figure}

When \(\hat{\zeta}_{m}=1\), we again find that \(\alpha\) decreases regardless of the sign of \(\mu\). For \(\mu>0\) then \(\Phi\) decreases and the behaviour is similar to the solutions found in section 5.1. The solution becomes dominated by the scalar field at late time and eventually hits a finite-time singularity, as shown in Figure 5.

By contrast when \(\mu<-2\) we find that now \(\Phi\) increases, and the behaviour is not so singular. Instead, we find that \(x_{1}\) quickly decreases and \(x_{3}\) slowly increases (Figure 6b). At late times it seems to dominate the dynamics completely.

\section{Conclusions}
In this paper we have studied a generalisation of the canonical BSBM theory of varying $\alpha =\exp [2\phi ]$, to include both a potential $V(\phi )$ and generalised coupling term $\omega (\phi )$ defining the theory. We were able to study in detail the dynamics of this model in FRW universes by formulating the equations as a dynamical system, and so study the full dynamics even in the cases where the background expansion is affected by the variations in $\phi $. This extends earlier studies of the situation with $\zeta _{m}<0$, which assumed that the variation of the $\phi $ field has negligible effects on the expansion scale factor of the universe and the evolution of the matter density. All the asymptotic behaviours were identified and some exact solutions found. We confirm the behaviour found in earlier approximate and numerical analyses in which the dynamics of $\phi$ were assumed not to affect the expansion dynamics of the universe to leading order \cite{barrow02, barrow08, barrow12}. We also studied the cases in which $\zeta _{m}>0$, whereas past studies have been confined to the situation with $\zeta _{m}<0$.

The general behaviour at early times is that the solution becomes dominated by the scalar field's kinetic energy, unless the potential is non-constant. The solution tends to eq. \eqref{3.13} as $t\rightarrow {0}$.

When only dust is present the late-time behaviour depends crucially on the sign of ${\zeta }_{m}$. In the zero curvature case, if ${\zeta }_{m}<0$ then the attractor is the Einstein-de Sitter solution, with $\alpha $ growing logarithmically in time. If ${\zeta}_{m}>0$ then after a transient Einstein-de Sitter phase, with decreasing $\alpha $, the dynamics become dominated by the scalar field and the solution develops a finite-time singularity. When a positive cosmological constant is added the late-time solution tends exponentially rapidly to a de Sitter universe with $\alpha $ frozen in at a constant value.

When dust is accompanied by negative curvature the solution at late times becomes curvature-dominated and evolves to a Milne universe with $\alpha $ frozen in at a constant value. The only different between this and the presence of a cosmological constant is that the evolution is slower in this case. With positive curvature the universe undergoes the process of collapse as in general relativity, although in the recollapsing phase the scalar-field kinetic energy dominates: $\alpha $ diverges like a power-law on approach to this singularity.

In the presence of radiation the late-time behaviour is the same as above. At early times, however, $\alpha $ may grow like a power-law of time depending on the initial conditions. We also showed that the behaviour of the radiation density and temperature in the presence of varying $\alpha $ is significantly different to the standard case where $\alpha $ cannot change: the radiation temperature falls as $T_{r}\varpropto \alpha^{1/4}a^{-1}$ and the radiation entropy per baryon is not constant during the expansion of the universe since $T_{r}^{3}/\rho _{m}\propto \alpha ^{3/4}$.

When the scalar field driving $\alpha $ variations has a non-zero self-interaction potential the evolution changes. For an exponential potential, $V(\phi )=\Lambda {}e^{\beta {}\phi }$, we find that when $\beta <0$ the late-time attractors are the power-law solutions \eqref{6.4} and \eqref{6.5}. In the less physically interesting case of $\beta >0$ the late-time behaviour may be a power-law attractor, or a finite-time singularity depending on the defining parameters. For a power-law potential, $V(\phi )=\Lambda \phi ^{n}$, again we find that when $n<0$ the late-time behaviour is a scalar field-dominated solution given by \eqref{6.11}.

We also investigated the effects of generalising the coupling parameter $\omega $ to become a function, $\omega (\phi )$. We studied the detailed evolution of the dynamics. With an exponential coupling, $\omega =\omega_{0}e^{\mu \phi }$, and a positive cosmological constant the late-time behaviour is, as before, the de Sitter solution. If the cosmological constant is dropped and the matter source is only dust then the late-time behaviour is more complex. When $\mu >-2$ and $\zeta _{m}<0$ the late-time attractor is the Einstein-de Sitter solution with $\alpha $ increasing as a power of a logarithm of time. When $\mu <-2$ and $\zeta_{m}>0$ then at late-times the scalar field dominates the dynamics and $\alpha $ decreases. When $\mu >-2$ and $\zeta _{m}>0$ or $\mu <-2$ and $\zeta _{m}<0$ finite-time singularities develop due to the scalar-field evolution \cite{finite}.

Our work has been strictly limited to homogeneous and isotropic universes. Given the recent experimental indications of a dipole in $\alpha $ it would be interesting to relax this assumption: in particular to examine small perturbations about this background (which were first studied for simple BSBM models with constant $\omega $ and $V=0$ using a gauge invariant formalism by Barrow and Mota \cite{barrow03}), or to look at larger deviations to determine the conditions under which modes for growing and decaying $\alpha(t)$ can coexist in the same solution. This will be explored elsewhere.

\section*{Acknowledgments}
A.A.H.G. and J.D.B. are supported by the STFC.

\appendix
\section{Dynamical systems analysis with closed curvature}
\begin{table} [t]
\begin{center}
  \begin{tabular}{| l || c | c | c | c | c | c | p{3.5cm} | }
    \hline
    SP & \(x_{0}\) & \(x_{H}\)  & \(x_{1}\) & \(x_{3}\) & Existence & Eigenvalues & Stability  \\ \hline
    \(1\) & \(1\) & \(1\) & \(0\) & \(0\) & all \(\omega\), \(\hat{\zeta}_{m}\) & \(0\), \(-\frac{3}{2}\), \(1\) & unstable saddle \\ 
    \(2\) & \(0\) & \(1\) & \(1\) & \(0\) & all \(\omega\), \(\hat{\zeta}_{m}\) & \(4\), \(3\), \(\frac{3}{2}-\sqrt{\frac{6}{\omega}}\) & unstable node \(\omega>\frac{8}{3}\), saddle point \(\omega<\frac{8}{3}\) \\ 
    \(3\) & \(0\) & \(1\) & \(-1\) & \(0\) & all \(\omega\), \(\hat{\zeta}_{m}\) & \(4\), \(3\), \(\frac{3}{2}+\sqrt{\frac{6}{\omega}}\) & unstable node \\ 
    \(4\) & \(0\) & \(1\) & \(\sqrt{\frac{8}{3\omega}}\) & \(\sqrt{\frac{3\omega-8}{3\omega}}\) & \(\omega>\frac{8}{3}\), \(\hat{\zeta}_{m}=-1\) &  \(1+\frac{8}{\omega}\), \(\frac{4}{\omega}-\frac{3}{2}\), \(\frac{8}{\omega}\) & saddle point \\ 
    \(5\) & \(1\) & \(-1\) & \(0\) & \(0\) & all \(\omega\), \(\hat{\zeta}_{m}\) & \(0\), \(\frac{3}{2}\), \(-1\) & unstable saddle \\ 
    \(6\) & \(0\) & \(-1\) & \(1\) & \(0\) & all \(\omega\), \(\hat{\zeta}_{m}\) & \(-4\), \(-3\), \(-\frac{3}{2}-\sqrt{\frac{6}{\omega}}\) & stable node \\ 
    \(7\) & \(0\) & \(-1\) & \(-1\) & \(0\) & all \(\omega\), \(\hat{\zeta}_{m}\) & \(-4\), \(-3\), \(-\frac{3}{2}+\sqrt{\frac{6}{\omega}}\) & stable node \(\omega>\frac{8}{3}\), saddle point \(\omega<\frac{8}{3}\) \\ 
    \(8\) & \(0\) & \(-1\) & \(-\sqrt{\frac{8}{3\omega}}\) & \(\sqrt{\frac{3\omega-8}{3\omega}}\) & \(\omega>\frac{8}{3}\), \(\hat{\zeta}_{m}=-1\) &  \(-(1+\frac{8}{\omega})\), \(\frac{3}{2}-\frac{4}{\omega}\), \(-\frac{8}{\omega}\) & saddle point \\ \hline
\end{tabular}
\end{center}
\caption{Stationary points for a universe with dust, scalar field and positive curvature. Points with \(x_{H}>0\) correspond to expanding universes, while \(x_{H}<0\) to collapsing universes. Point 6 is the global attractor for this system.}
\end{table} 

To perform a dynamical systems analysis for a closed universe the formalism of section 4 is not ideal, since $H\rightarrow {0}$ if there is a point of maximum expansion and the variables \eqref{4.1} diverge on approach to it. To remedy this we will follow the strategy of \cite{goliath99}. We define 
\begin{equation} \label{a.1}
D=\sqrt{H^{2}+k/a^{2}}, 
\end{equation}
which is finite at the turnover, and use it to define the new variables
\begin{equation} \label{a.2}
x_{1}=\frac{\sqrt{\omega}\dot{\phi}}{\sqrt{6}D},\ x_{2}=\frac{\sqrt{V}}{\sqrt{3}D},\ x_{3}=\frac{\sqrt{\rho_{m}|\zeta_{m}|}e^{-\phi}}{\sqrt{3}D},\ x_{4}=\frac{\sqrt{\rho_{r}}e^{-\phi}}{\sqrt{3}D},\ x_{H}=\frac{H}{D}. 
\end{equation}
The curvature term \(x_{5}=\sqrt{|k|}/aD\) is not needed since by \eqref{a.1} \(x_{5}^{2}=1-x_{H}^{2}\). As before we also have that \(x_{0}=\sqrt{\rho_{m}/3D^{2}}=1-x_{1}^{2}-x_{2}^{2}-x_{3}^{2}-x_{4}^{2}\). We will define a modified time coordinate
\begin{equation} \label{a.3}
\mathcal{N}=\int{D}dt. 
\end{equation}
Using this the autonomous variables obey the evolution equations
\begin{align}
&\frac{dx_{H}}{d\mathcal{N}}=-\frac{1}{2}(1-x_{H}^{2})(1+3x_{1}^{2}-3x_{2}^{2}+x_{4}^{2}), \label{a.4}
\\
&\frac{dx_{1}}{d\mathcal{N}}=-\sqrt{\frac{6}{\omega}}\hat{\zeta}_{m}x_{3}^{2}+\sqrt{\frac{3}{2\omega}}\lambda{}x_{2}^{2}-\frac{1}{2}x_{1}x_{H}(3-3x_{1}^{2}+3x_{2}^{2}-x_{4}^{2}), \label{a.5}
\\
&\frac{dx_{2}}{d\mathcal{N}}=-\sqrt{\frac{3}{2\omega}}x_{1}x_{2}\lambda+\frac{1}{2}x_{2}x_{H}(3+3x_{1}^{2}-3x_{2}^{2}+x_{4}^{2}), \label{a.6}
\\
&\frac{dx_{3}}{d\mathcal{N}}=-\sqrt{\frac{6}{\omega}}x_{1}x_{3}+\frac{1}{2}x_{3}x_{H}(3x_{1}^{2}-3x_{2}^{2}+x_{4}^{2}), \label{a.7}
\\
&\frac{dx_{4}}{d\mathcal{N}}=-\frac{1}{2}x_{4}x_{H}(1-3x_{1}^{2}+3x_{2}^{2}-x_{4}^{2}). \label{a.8}
\end{align}
When \(x_{H}>0\) the solution is expanding, while when \(x_{H}<0\) it is collapsing. The point of maximum expansion occurs when \(x_{H}=0\). If we restrict to universes containing just dust and curvature then we see that there are 8 stationary points given in Table 7. For each stationary point on the expanding branch there is a corresponding one on the contracting branch, but attractors only exist on the contracting branch. This tells us these solutions always recollapse as in general relativity. The only difference is that the solution becomes dominated by the kinetic energy of the scalar field in the final stages of collapse. Near the point of collapse \(t_{0}\) the dynamics are given by
\begin{equation} \label{a.9}
a(t)=A(t_{0}-t)^{\frac{1}{3}},\ {}\ \phi(t)=\pm\sqrt{\frac{2}{3\omega}}\ln(t_{0}-t)+\phi_{0}, 
\end{equation}
where \(A\) and \(\phi_{0}\) are constants. This means \(\alpha\) behaves like a power-law, \(\alpha=\alpha_{0}(t_{0}-t)^{\pm\sqrt{\frac{8}{3\omega}}}\), near the collapse. This is confirmed by numerical simulations of the system. One finds that for \(\zeta_{m}>0\) the behaviour of \(\alpha\) is described by the solution with the \(+\) sign (so \(\alpha\rightarrow{0}\)), while for \(\zeta_{m}<0\) it is the \(-\) sign (so \(\alpha\rightarrow{\infty}\) at the 'big crunch' singularity).

\section{Solutions with scalar-field domination}
Here we show that it is possible to solve the equations \eqref{3.2}-\eqref{3.3} exactly in the limit where one of the scalar terms dominates the Friedmann equation. For simplicity let us restrict to zero potential and constant coupling (these assumptions can be relaxed somewhat). We shall also ignore the effects of radiation, and set \(k=\Lambda=0\). The first approximation means we are restricting to solutions valid at late times; the second assumption is because we know from the analysis of section 5 that otherwise the solutions will never become scalar dominated at late time. We wish to solve the equations
\begin{align} 
&\frac{\dot{a}^2}{a^2}=\frac{1}{3}\left(\rho_{m}(1+|\zeta_{m}|e^{-2\phi})+\frac{1}{2}\omega\dot{\phi}^2\right), \label{b.1}
\\
&\ddot{\phi}+3H\dot{\phi}=N\frac{e^{-2\phi}}{a^{3}}, \label{b.2}
\end{align}
where \(N=-2\zeta_{m}\rho_{m}a^{3}/\omega\) is a constant The first limit we shall look at is when the kinetic energy of the scalar field dominates over the other terms in the Friedmann equation. This may be the late-time limit of the dust solutions with exponential coupling, \(\zeta_{m}<0\) and \(\mu<-2\) considered in section 7.1. In this limit the Friedmann equation becomes \(\frac{\dot{a}}{a}\approx\sqrt{\frac{\omega}{6}}\dot{\phi}\). This may be integrated to yield
\begin{equation} \label{b.3}
a=a_{0}e^{\sqrt{\frac{\omega}{6}}\phi}. 
\end{equation}
Substituting this into the scalar equation gives us that
\begin{equation} \label{b.4}
\ddot{\phi}+\sqrt{\frac{3\omega}{2}}\dot{\phi}^{2}=N'e^{-\left(2+\sqrt{\frac{3\omega}{2}}\right)\phi}.
\end{equation}
Since this equation does not depend on \(t\) explicitly it may be reduced to a 1st order equation. In fact, putting \(u(\phi)=\dot{\phi}^{2}\) and replacing \(t\) with \(\phi\) gives us a linear equation in \(u(\phi)\),
\begin{equation} \label{b.5}
\frac{du}{d\phi}+\sqrt{6\omega}u=2N'e^{-\left(2+\sqrt{\frac{3\omega}{2}}\right)\phi}. 
\end{equation}
It is worth noting that similar steps would give a solvable equation even when \(\omega=\omega(\phi)\). Integrating this gives 
\begin{align} 
&u(\phi)=\dot{\phi}^{2}=\left(Ce^{-\sqrt{6\omega}\phi}+2N'\left(\sqrt{\frac{3\omega}{2}}-2\right)^{-1}e^{-\left(2+\sqrt{\frac{3\omega}{2}}\right)\phi}\right), \label{b.6}
\\
&\implies{}t=\bigintsss{}\frac{d\phi}{\sqrt{\left(Ce^{-\sqrt{6\omega}\phi}+2N'\left(\sqrt{\frac{3\omega}{2}}-2\right)^{-1}e^{-\left(2+\sqrt{\frac{3\omega}{2}}\right)\phi}\right)}}+t_{0}, \label{b.7}
\end{align}
where \(C\) and \(t_{0}\) are constants. Equations \eqref{b.3} and \eqref{b.7} constitute the parametric solution for the scale factor in this limit. The above integral can only be done in general with the aid of hypergeometric functions, though it does simplify when \(C=0\). In this case there is a power-law solution of the form
\begin{equation} \label{b.8}
\phi=A+B\ln{t},\ {}\ a(t)=a_{0}t^{\sqrt{\frac{\omega}{6}}B},
\end{equation}
where \(A\) and \(B\) are constants given by
\begin{equation} \label{b.9}
B=\frac{1}{1+\sqrt{3\omega/8}},\ {}\ A=\frac{1}{2}\ln{}\left(\frac{N'(\sqrt{3\omega/8}+1)^{2}}{(\sqrt{3\omega/8}-1)}\right).
\end{equation}
Note these particular solutions cannot be the attractors if the scalar field does indeed dominate in this manner. We can easily see this by noting they are not consistent solutions: the \(\rho_{m}\) and \(\rho_{m}e^{-2\phi}\) terms always decay slower in the Friedmann equation than the kinetic term \(\frac{1}{2}\omega\dot{\phi}^{2}\).

The other limit in which we can solve these equations is when the 2nd term in \eqref{b.1} dominates over all others. This seems to be the late-time behaviour of the dust solutions with \(\zeta_{m}>0\) and \(\mu<-2\) discussed in section 7.1. In this limit \(H^{2}\approx{}\lambda^{2}e^{-2\phi}/a^{3}\) where \(\lambda^{2}=\rho_{m}a^{3}|\zeta_{m}|/3\) is another constant. Substituting this into \eqref{b.2} gives us 
\begin{equation} \label{b.10}
\ddot{\phi}+\frac{3\lambda{}e^{-\phi}}{a^{3/2}}\dot{\phi}=\frac{Ne^{-2\phi}}{a^{3}}. 
\end{equation}
If we now replace derivatives of \(t\) with \(a\) then it becomes (writing \('=d/da\) from now on)
\begin{equation} \label{b.11}
a^{2}\phi''+\frac{5}{2}a\phi'-a^{2}\phi'^{2}=\frac{N}{\lambda^{2}}. 
\end{equation}
Since this equation does not depend on \(\phi\) explicitly it may be reduced to a 1st order equation by putting \(w(a)=a\phi'\) to give
\begin{equation} \label{b.12}
aw'=w^{2}-\frac{3}{2}w+\frac{N}{\lambda^{2}}, 
\end{equation}
which is a separable equation. Notice that this equation admits simple particular solutions corresponding to \(w=w_{0}\), where \(w_{0}\) is a constant given by
\begin{equation} \label{b.13}
w_{0}=\frac{3}{4}\pm\frac{1}{2}\sqrt{\frac{9}{4}-\frac{4N}{\lambda^{2}}}. 
\end{equation}
Since \(\phi'=w/a\) then in this case \(\phi(a)=w_{0}\ln{a}+\phi_{0}\). Substituting this into the Friedmann equation shows that these are power-law solutions with 
\begin{equation} \label{b.14}
a(t)=a_{0}t^{\frac{2}{3+2w_{0}}}. 
\end{equation}
Note that these solutions only exist when \(\delta^{2}>0\), but since \(N/\lambda^{2}=-\frac{6}{\omega}\hat{\zeta}_{m}\) this is always satisfied for \(\zeta_{m}>0\). In this case then the larger and smaller value of \(w_{0}\) are positive and negative respectively. 

Since \eqref{b.12} is separable, we can find the general solution by integration. This gives
\begin{equation} \label{b.15}
a=a_{0}\left(\frac{\delta+3/2-2w}{2w-3/2+\delta}\right)^{1/\delta} \mbox{ with }\ {}\ \delta=\sqrt{\frac{9}{4}-\frac{4N}{\lambda^{2}}}.
\end{equation}
This can be inverted to give \(w(a)\)
\begin{equation} \label{b.16}
w(a)=\frac{3}{4}+\frac{\delta}{2}\left(\frac{a_{0}^{\delta}-a^{\delta}}{a_{0}^{\delta}+a^{\delta}}\right), 
\end{equation}
so that the scalar field is then 
\begin{equation} \label{b.17}
\phi(a)=\int\frac{w(a)}{a}da=\left(\frac{3}{4}+\frac{\delta}{2}\right)\ln{}a-\ln{}(a_{0}^{\delta}+a^{\delta}). 
\end{equation}
This allows us to find the solution implicitly, since if we substitute this back into the Friedmann equation the time is given by
\begin{equation} \label{b.18}
t=\int\frac{a^{1/2}e^{\phi}}{\lambda}da=\frac{1}{\lambda}\int{}\frac{a^{5/4+\delta/2}}{a_{0}^{\delta}+a^{\delta}}da. 
\end{equation}
This gives the implicit solution for the scale factor and the scalar field.

We have not been able to solve these equations analytically in the limit when both scalar terms in the Friedmann equation are non-negligible. There is one case of this kind which can be solved though. If we know that at late times these two terms approach some ratio (i.e. that \(x_{1}/x_{3}\) tends to a constant) then we can use either of these solutions given above to find the general solution.

\end{document}